\pdfoutput=1
\documentclass[lettersize,journal]{IEEEtran}
\usepackage{amsmath,amsfonts}
\usepackage{algorithm}
\usepackage{array}
\usepackage[caption=false,font=normalsize,labelfont=sf,textfont=sf]{subfig}
\usepackage{textcomp}
\usepackage{stfloats}
\usepackage{url}
\usepackage{verbatim}
\usepackage{graphicx}
\usepackage{cite}
\usepackage{graphicx} 
\usepackage{caption} 
\usepackage[dvipsnames]{xcolor}
\usepackage{comment}
\usepackage{amsmath} 
\usepackage{textalpha}
\usepackage{tabularx} 
\usepackage{multirow}
\usepackage{graphicx}
\usepackage{array}
\usepackage{tabularx}
\usepackage{calc} 
\usepackage{amssymb}  
\usepackage{booktabs}
\usepackage{orcidlink}
\usepackage{balance}

\usepackage{algorithm}
\usepackage[noend]{algpseudocode} 

\usepackage{enumitem}
\SetEnumitemKey{algfix}{label=\arabic*.,leftmargin=*,align=left}

\algrenewcommand\alglinenumber[1]{\footnotesize #1}
\algrenewcommand\algorithmiccomment[1]{\hfill\(\triangleright\)~#1}
\algrenewcommand\textproc{\textsc} 

\makeatletter
\renewcommand{\ALG@name}{Algorithm} 
\makeatother

\usepackage{hyperref}

\hypersetup{%
  colorlinks=true,%
  linkcolor={blue},
  citecolor={blue},
  urlcolor={blue!80!black},
  bookmarksnumbered=true,%
  bookmarksopen=true}

\makeatletter
\def\@cite#1#2{{\color{blue}[#1]\if@tempswa , #2\fi}}
\makeatother

\begin{document}

\title{Unlearning to Protect: A Distilled Reinforcement Learning Framework with Privacy-Preserving Feature Unlearning and XAI for IoT Security}


\author{Md. Nahid Hasan\textsuperscript{\orcidlink{0009-0005-8099-0650}} and Md. Golam Rabiul Alam\textsuperscript{\orcidlink{0000-0002-9054-7557}}\\ \normalsize Department of Computer Science and Engineering, BRAC University}





\maketitle

\begin{abstract}
Botnets pose a significant cybersecurity threat, enabling attacks such as DDoS, data theft, and service disruptions on IoT devices. These devices often lack built-in botnet traffic filtering, leaving them highly exposed. Existing AI-based solutions improve detection capabilities but have limitations: (i) they are too heavy for IoT deployment, and (ii) they lack unlearning capabilities to forget sensitive or outdated features without retraining. To address these challenges, we propose DiRLU, a lightweight, reinforcement learning driven framework, while ensuring privacy by selectively unlearning sensitive or outdated features without requiring retraining. The framework leverages knowledge distillation to transfer knowledge from a teacher model into a lightweight student model, with both models trained using A2C. A post-hoc unlearning mechanism modifies weights to remove targeted features, while restored features show negligible performance loss, confirming reversibility. Unlike many benchmark models that used only 5\% of the BoT-IoT dataset, this research leverages 25\%, allowing us to develop a strong teacher model. Both the teacher and student models were trained using the A2C reinforcement learning algorithm, achieving impressive results, with the student model achieving 99.60\% accuracy and a 99.80\% F1 score. To enhance transparency, we integrated Explainable AI (XAI), particularly LIME, which helps interpret the model’s decisions and identify the key features influencing its predictions. Moreover, DiRLU requires only 2,370 FLOPS, approximately 3.87$\times$ more efficient than the state-of-the-art model, highlighting its efficiency for edge deployment. DiRLU combines efficiency with privacy, aligning with GDPR standards (right to be forgotten) to provide practical and scalable IoT security solution.
\end{abstract}


\section{Introduction}
\IEEEPARstart{R}{obot} network, in short, Botnet, is a computer network infected by malware under the control of an attacking party known as Bot Herder \cite{n1}. Computers that are under the control of attackers are called bots. Attackers having remote access to computers through bots can read, update, and even decrypt sensitive data for financial gain or reputational loss. Controlling compromised devices over the networks remotely is a vital part for attackers. Bots are controlled directly or indirectly in two modes: Centralized client-server models and Decentralized peer-to-peer models. In 2016, Mirai botnet attack was on Dyn, an Internet performance management company. During the outage, the estimated losses were 22,000 US dollars per minute \cite{n2}. Modern malware, ransomware, and botnets pose serious cybersecurity risks to individuals, businesses, and governments.

Malicious content implants individuals' or organizations' devices in diverse ways, often exploiting human error and system vulnerabilities. Attackers typically start with phishing emails or websites, misleading victims into clicking malicious hyperlinks or downloading infected attachments. Once inside, botnets can spread across networks, gaining control of devices to carry out intended vicious activities. These infected systems are then used to deploy ransomware. Ransomware is malicious software or program that encrypts a victim's data, making it inaccessible until a ransom is paid to the attacker. Nevertheless, ransomware attacks can lead to a significant service outage for an organization. Attackers send commands to bots for launching attacks such as DDoS, Reconnaissance, Information Theft, Service Scan, Keylogging, etc.

When cyber threats such as DDoS, reconnaissance, information theft, or service scanning infiltrate an individual's or organization's security, they can impose significant damage over time if not noticed promptly. A distributed denial of service (DDoS) attack can immediately hinder services, rendering websites, vital tools, or devices unreachable and halting routine operations. Reconnaissance helps attackers to traverse the local network and discover weak and sensitive systems. Again, Information theft targets data such as client information or credentials, posing a risk and incurring reputational damage. Another type of attack, service scans, assist attackers in locating entry points into an organization's systems, which they then exploit to spread malware or secure unauthorized access. Keylogging secretly captures somebody's keystrokes, capturing confidential information and letting attackers control accounts or steal confidential data. Over time, these attacks seize control, disrupt operations, and expose the organization to wider breaches.

\subsubsection{Contribution}
To protect an individual's or organization's sensitive data or reputation, we have to build and implement such security that can be adopted by any network device and is capable of defending against attacks before any botnet beaches inside. In this research, we developed a customized reinforcement learning based compact model named DiRLU that can detect and identify botnet attacks in the network layer. The model was trained on a big dataset where attack network traffic dominated normal traffic, which mimics realistic cyberattack scenarios in real-time. Further, to have a trustworthy model, we employed  XAI to explain the model's inside mechanism. Additionally, we demonstrated the feature un-learning process to maintain data security \& privacy. Figure-\ref{fig0} represents the high-level architecture of the research. This research's influential and significant contributions are summarized as follows:

\begin{figure}[htbp]
  \centering
  \includegraphics[width=.97\columnwidth]{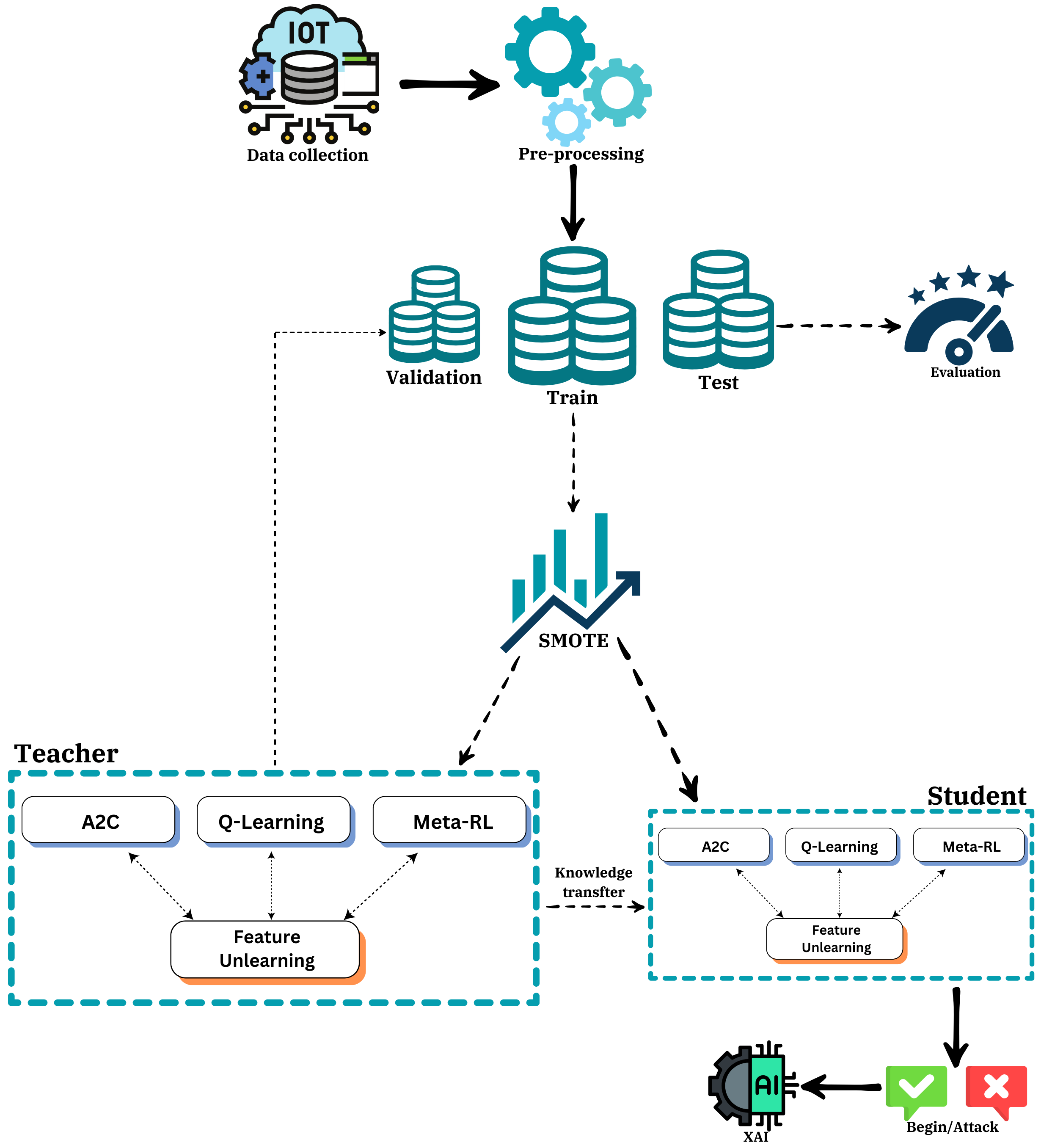}
  \caption{A high-level overview of the proposed research}
  \label{fig0}
\end{figure}

\begin{itemize}
    \item We propose DiRLU, a lightweight yet effective reinforcement learning framework using knowledge distillation to transfer knowledge from a large teacher model to a smaller student model. The student model is optimized with fewer hyperparameters, making it suitable for deployment on edge devices, such as mobile phones, surveillance cameras, and security gateways.
    
    \item To maintain user privacy and ensure model adaptability, we implement feature unlearning to remove sensitive (features that dominate or contribute most to the model’s predictions) or outdated attributes from the trained model without full retraining. This promotes GDPR compliance \cite{GDPR} and makes the model scalable.

    \item We reintroduce unlearned features to validate the effectiveness of the unlearning process. This comparison helps evaluate the impact of feature removal on model performance, ensuring the system remains robust and extendable.
    
    \item While benchmark models typically use only 5\% of the dataset, we utilize a more diverse and representative 25\% subset of the Bot-IoT dataset. This ensures better class coverage, improved generalization, and more reliable detection performance, especially in handling rare attack patterns.
    
    \item Explainable AI (XAI) techniques, such as LIME, are integrated to visualize and interpret model predictions. This enhances trust by demonstrating the reasoning behind attack classifications and helping cybersecurity analysts in decision-making.
\end{itemize}

The rest of the paper has been organized as follows: Section-\hyperref[related_work]{II} explains some relevant research along with their contributions and limitations. Section-\hyperref[methodology]{III} describes the methodology, dataset and algorithms used in this research in detail. Section-\hyperref[result_and_analysis]{IV} represents the results we obtained from this study, including XAI, to ensure the transparency of this research. We also describe some limitations and future goals of this paper. Finally, in Section-\hyperref[conclusion]{V}, we included a consultation, providing the study's findings and potential areas for further exploration.

\section{Related Works}\label{related_work} 
Several related studies have already been conducted to detect Botnet network traffic. This section briefly summarizes some studies and their approaches to detecting Botnet network traffic.

Leevy et al. \cite{n3} build a successful predictive machine learning model for detecting information theft attacks. This study used eight ML classifiers (four ensembles and four non-ensembles) and two evaluation metrics (AUC and AUPRC). The authors used only six features of the Bot-IoT dataset to get only specific insights about information theft attacks. The authors used 556 instances and other cutting-edge technologies, such as k-fold cross-validation, for training the model. Based on AUC and AUPRC results, the ensemble classifiers performed better than the non-ensemble classifiers. However, the LightGBM ensemble classifier performed top among all; it exhibited AUC results of 98\% and AUPRC of 99\%.

Koroniotis et al. \cite{n4} proposed a dataset, Bot-IOT, to simulate different types of network attacks, such as probing attacks, DDoS, DoS, and Information theft. Authors build the dataset to address other datasets' drawbacks with accurate labeling and incorporate recent attacks. They captured one GByte raw packet to extract complete information on network features. They introduced 32 features and created ten other hybrid features for simulation and testing the dataset's accuracy. They also evaluated the reliability of the dataset using ML (SVM) and deep learning algorithms (RNN and LSTM). SVM performed best with a full-featured dataset and scored 99\% accurately, whereas RNN and LSTM performed 99\% accurately with the ten best features only.

Alosaimi et al. \cite{n5} developed a machine learning-based intrusion detection system to secure IOT applications. The authors used the BOT-IoT dataset and a combination of deep learning and machine learning algorithms to train and test the model. The authors only used four files of the dataset, containing only 5-6\% of the dataset. The first file contains only a DoS attack, the second one includes a DoS and DDoS attack, The third one contains two DDoS attacks, and The fourth one contains many attacks. They evaluated the results using accuracy, error rate, recall, and specificity. However, in the second database, the ensemble bag algorithm acquired an accuracy of 100\% at all levels of the database, while the decision tree obtained an accuracy of 100\% at all levels of the database in the third database. This study demonstrates remarkable improvements in intrusion detection, especially DoS attacks, compared to existing models.

Farrokhmanesh et al. \cite{n6} designed a new machine learning-based model for detecting malicious activity in computers and networks. The authors proposed their model, which is based on audio signal processing techniques. They represented binary bytes as audio signals and applied MIR techniques for detecting malware. The authors used 3400 samples containing both malware and benign, however, they used the VX Heavens dataset for malware and Windows XP files for benign. They used KNN, AdaBoost, and Random Forest classifiers, among them AdaBoost performed best in terms of accuracy (92.2\%). The authors demonstrated a lightweight novel model with good accuracy and is computationally efficient.

Bensaoud et al. \cite{n7} arranged a survey to examine how deep learning (DL) detects malware across multiple operating systems, including MacOS, Windows, iOS, Android, and Linux. They investigate different Deep Learning techniques, including text and image classification, pre-trained models, and multi-task learning, to determine how well they detect malware software. The paper also highlights significant problems, such as defined benchmark datasets, the complexity of deep learning models, and their sensitivity to adversarial attacks. It also emphasizes the value of Explainable AI (XAI) in making these models more transparent and intelligible. This study provides valuable insights into malware detection techniques by testing eight deep-learning approaches on various datasets.

Wang et al. \cite{n8} study machine unlearning, which refers to a trained model's ability to "forget" or erase some information. This survey concentrated on type of methods, from easy ones like deleting data and retraining, to more advanced one which is gradient-based methods that modify the model without retraining the model as a whole. They present several techniques and give an order to them depending on their performance, cost in terms of computing power, and their impact on the accuracy of the model. The research paper demonstrates that it is very difficult to apply unlearning forever without affecting general performance of the learning system. It points out shortcomings in current works that require benchmarking for consistency among different researchers. The paper posits potential ways forward for increasing the scalability and utility value of unlearning in large-scale machine-learning systems.

Miao et al. \cite{n9} in DistillMal introduce a novel method to teach the smaller studentship more effectively using knowledge distillation. The rationale behind this is to train a small student network as if it were a large and complex teacher network. This resulted in a compact model that can be used on resource-constrained devices. We did plenty of tests on two different additional datasets and found out that student network functions almost as good as the teacher model. Detection accuracy of the student model was 94.5\%, while detection accuracy of the instructor’s model was 95\%. Besides, the number in the student model reduced by 60\%, and the time consumption for making inference decreased by 50\% compared to the instructor model. This is an indication that DistillMal is able to achieve a reasonable performance trading off between model size and detection accuracy. The proposed approach uses knowledge distillation for addressing issues related to the applicability of deep learning models, which are large-scale ones within low computing power scenarios, like mobile devices or IoT systems.

In their research on the “Explainability in AI behavioral malware detection systems”, Galli et al. \cite{n10} infer that it is caused by the growing need for transparency in machine learning models that find patterns of BEHAVIOR to detect malevolent software. In order to avoid this, they created a model of detection which provides explanations through SHapley Additive exPlanations (SHAP) on how every feature affects a decisional process of ML approaches. Cybersecurity professionals are thus given an opportunity to judge what types of behavioral signs should be weighted more while assessing malware or benign software. Their approach was highly accurate as evident from its performance metric when it comes to the task of detecting malware. The solution not only possesses explainability along with detection capability for increasing threat response, it also helps to build trust and comply with regulatory criteria as well. Thus, the study highlights that there is no necessity either for very strong detection techniques or for very complex AI-based insight solutions that could be easily grasped by end-user system staff.

EDIMA was proposed by Kumar et al. \cite{n11} as a solution to the problem of detecting IoT botnets at the early stages of their infection on computers in home networks. EDIMA merges the classification of traffic with the application of Autocorrelation Function (ACF) for the detection of possible bots’ behavior patterns. Their findings indicate that EDIMA is very effective in correctly identifying both CnC and bot scanning activities hence recording a 98\% detection rate whilst suffering fewer false alarms. The authors explain that this performance stems directly from the increase in the number of IoT devices, with his or her work being benchmarked on the use of Raspberry Pi where they indicated short delays in detection as well as low resource usage on EDIMA part. This technology surpasses all other prevailing solutions that deal with identification and prevention/ mitigation against botnet’s threats on real-time Skype video streaming scenario-DzMulla and Zwiggelaar, 2012 [1]. Moreover, even though AI is implemented for real-time operation through this technology in the edge processor does not bring about any processing delay appreciably. The methodology is an effective measure to guard the IoT environment from bot activities efficiently and within manageable limits system has been developed.

The authors introduce a paper that tackles a problem, "PAIRED: An Explainable Lightweight Android Malware Detection System," further describes PAIRED \cite{n12}. To detect malware programs, PAIRED utilizes machine learning as well as static analysis methods. The system's capability to identify malicious from harmless programs is shown by its 94\% detection accuracy. In addition, PAIRED is very suitable for use in devices with little computational power, thanks to its lightweight architecture that makes it not resource-intensive. Through AI algorithms, which are explainable consumers can understand what features constitute a detection decision. One of the possible solutions for such threats caused by the increased usage of Android smartphones is represented by PAIRED, which combines both explainability and high detection rates.

In their exploration of Large Language Models (LLMs), Patsakis et al. \cite{n13} explored the use of LLMs to deobfuscate malware with a special focus on the Emotet malware campaign. Their research employed four different LLMs to check whether these models are able to interpret and simplify obfuscated code. The models had trouble with the most complex types of obfuscation, but they were capable of precisely recovering portions of hazardous payloads, obtaining accuracy rates as high as 85\%. These results suggest that LLMs may become a significant asset in the field of malware study; however, further improvements should be made to suitably solve current security issues.

Wu et al. \cite{n14} presented an architecture of privacy-preserved deep learning for botnet malware detection that does not allow privacy to be leaked. Rather than using the proposed method, which masks sensitive information in network traffic without compromising on effective features used for distinguishing botnet behaviors. To accomplish this, they have proposed a privacy-adversarial technique depending on mutual information with respect to minimizing the connection between personal data and anonymized features. Their framework introduces federated learning, which is a training process for a feature extractor through multiple users, which further enhances decentralization of data and thus its confidentiality. For perceptibility purposes, the functionality of this system involves converting network traffic into grayscale images and applying Self-Attention Networks along with CNNs and Bi-LSTM architectures in series/sequence. The outcomes when evaluated over ISCX-2014 as well as CTU-13 datasets demonstrated excellent performance: 100\% accuracy as well as F1-score in CTU-13; whereas without federated learning got 98.44\% accuracy with a 98.95\% F1-score in ISCX-2014 dataset. With the introduction of federated learning, there was a slight drop in accuracy (96.88\% accuracy and 97.92\% F1), but there was significant improvement in the privacy-preservation aspect . In comparison, the framework had higher efficiency than other latest techniques without compromising the strong safeguard of private details and a high identification level system.

Uprety et al. \cite{n15} provide a comprehensive survey on how reinforcement learning (RL) and deep reinforcement learning (DRL) can be used to improve the security of IoT. The paper singles out main IoT threats, such as DoS or DDoS attacks, jamming, and spoofing, for detailed decomposition and depicts how methods based on RL can handle each case. On the one hand, this may include multi-agent RL for protection against DDoS attacks; on the other hand, DRL for jamming defense power control; and, further, Q-learning for spoofing detection. In cyber-physical systems such as smart grids and intelligent transportation systems, attackers and defenders can be modeled using RL, enabling real threats to be detected early enough to prevent harm. The survey results indicate that techniques based on RL or DRL are not only more flexible but also faster to deploy and more resilient in protecting a system than some conventional approaches. Still, challenges remain, such as processing high-dimensional data, limited observability, cooperation among many agents, coordination, and defending against attacks targeting an RL system itself. The authors point out that more work will be needed to scale these methods for general adoption in community practice.

Muhatti et al. \cite{n16} describe an AI-based system for deep reinforcement learning network intrusion detection systems and introduce NIDS as the application of Asynchronous Advantage Actor-Critic (A3C). They argue that many AI-driven NIDS present challenges since, based on the fact that there are no efficient ways to collect or evaluate network data, they are not powerful enough to address new threats. Towards correction of this shortcoming, they leveraged automated network scanning and OSS technologies for improved data acquisition. The core objective is the development of A3C based NIDS that has the ability to detect known threats and zero-day attacks in dynamic environments by integrating value prediction with policy learning. This is verified on three benchmark datasets running benign requests as opposed to fake poisoning attempts. System with 300,000 nodes behaves like reality would do, thus, this method can be considered accurate (98.68\%) and has significantly fewer false alarms than other methods. Hence, this model is consistent and robust even in different scenarios, confirming that an anomaly-based approach could be a good one towards a dynamic threat protection system.

A number of studies have been conducted to address the issue of how to use deep learning and AI tools in classifying the botnet traffic effectively. Unlike Table-\ref{lit_summary}, which gives a summary of these studies and therefore provides their strengths and weaknesses.

\renewcommand{\arraystretch}{1.2} 
\begin{table*}[!ht]
    \caption{Summary of different studies}
    \centering
    \begin{tabular}{
        >{\raggedright\arraybackslash}m{0.5cm} 
        >{\raggedright\arraybackslash}m{2.5cm} 
        >{\raggedright\arraybackslash}m{2.5cm} 
        >{\raggedright\arraybackslash}m{3.6cm} 
        >{\raggedright\arraybackslash}m{1.2cm} 
        >{\raggedright\arraybackslash}m{1.2cm}
        >{\raggedright\arraybackslash}m{3.6cm}
    }
    \hline \hline
    \multicolumn{1}{c}{\textbf{Ref.}} &
    \multicolumn{1}{c}{\textbf{Dataset}} &
    \multicolumn{1}{c}{\textbf{Feature}} &
    \multicolumn{1}{c}{\textbf{Model}} &
    \multicolumn{1}{c}{\textbf{Unlearning}} &
    \multicolumn{1}{c}{\textbf{KD}} &
    \multicolumn{1}{c}{\textbf{Accuracy}}  \\ \hline

    \cite{n3} & Bot-IoT (5\%; Normal \& Information Theft classes) & 37 features & CatBoost, LightGBM, XGBoost, Random Forest, Decision Tree, Logistic Regression, Naive Bayes, MLP & N/A & N/A & Best: LightGBM AUC=0.987, AUPRC=0.994 ($\approx$98.34\% accuracy) \\

    \cite{n4} & Bot-IoT (5\% subset) & Original \& generated (10-best features selected via correlation \& joint entropy analysis) & SVM, RNN, LSTM & N/A & N/A & SVM (10-best): 88.37\%, \newline SVM (full): 99.99\%, \newline RNN (10-best/full): 99.74\%, \newline LSTM (10-best/full): 99.74\% \\

    \cite{n5} & Bot-IoT & 35 features & Decision Tree, Ensemble Bag, KNN, Linear Discriminant, SVM & N/A & N/A & DT: 99.99\%,\newline Ensemble Bag: 100\%,\newline KNN: 99.98\%,\newline SVM: 99.99\%,\newline LD: 100\% \\

    \cite{n6} & Custom Malware/Benign Dataset (3400 samples; PE and non-PE) & 34 features (26 from MFCC + Chromagram) & KNN, AdaBoost, Random Forest & N/A & N/A & KNN: 90.0\%, \newline AdaBoost: 92.2\%, \newline Random Forest: 92.0\% \\

    \cite{n7} & Malimg Dataset (grayscale/RGB malware images) & Grayscale and RGB images (converted from malware binaries) & EfficientNet, Inception V4, Xception, CapsNet & N/A & N/A & EffNet: 95.64\%, \newline Inception V4: 95.98\%, \newline Xception: 89.50\%, \newline CapsNet: 88.64\% \\

    \cite{n9} & VirusShare & 512 features (API sequence length) & MLP, TextCNN, Catak (BiLSTM), BERT-base, DistillBert, DistillMal & N/A & Teacher: BERT \newline Student: TextCNN & MLP: 83.5\% \newline TextCNN: 88.4\% \newline Catak (BiLSTM): 66.3\% \newline BERT-base: 89.2\% \newline DistillBert: 89.0\% \newline DistillMal: 89.1\% \\

    \cite{n10} & Mal-API-2019 & 128 (sequence length) & LSTM, BiLSTM, GRU, Attention, MultiHeadAttention & N/A & N/A & LSTM: 47.53\%, \newline BiLSTM: 52.88\%, \newline GRU: 47.69\%, \newline Attention: 52.18\%, \newline MultiHeadAttention: 47.69\% \\

    \cite{n11} & IoT-BPR-NSS Testbed \& UNSW IoT & 6 & Random Forest, SVM, Gaussian Naive Bayes & N/A & N/A & \textbf{Testbed:} RForest: 100\%, SVM: 99\%, GNB: 97\% \newline \textbf{UNSW IoT:} RForest: 100\%, SVM: 91\%, GNB: 92\% \\

    \cite{n12} & Drebin-215, Malgenome-215 \& CICMalDroid2020 & 35 & Random Forest, Logistic Regression, Decision Tree, Gaussian Naive Bayes, SVM & N/A & N/A & RF: Drebin-215: 98.07\%, Malgenome-215: 98.73\%, CICMalDroid2020: 97.98\% \\

    \cite{n13} & Emotet malware campaign (2,000 obfuscated PowerShell scripts; 2,869 URLs, 2,512 domains) & N/A (LLM prompt-based extraction) & GPT-4, Gemini Pro, Code Llama 34B Instruct, Mixtral 8x7B & N/A & N/A & \textbf{URL extraction accuracy}: GPT-4: 69.56\%, Gemini Pro: 36.84\%, Code Llama: 22.13\%, Mixtral: 11.59\%; \newline \textbf{Domain extraction accuracy}: GPT-4: 88.78\%, Gemini Pro: 55.99\%, Code Llama: 35.46\%, Mixtral: 24.92\% \\

    \cite{n14} & ISCX-2014 \& CTU-13 botnet traffic datasets & Raw traffic bytes (image-based input) & CNN + Bi-LSTM with privacy-adversarial feature extractor trained via mutual information minimization \& federated learning & N/A & N/A & \textbf{CTU-13}: 100\% accuracy/F1 (both FL and non-FL) \newline \textbf{ISCX-2014}: non-FL: Acc=98.44\%, F1=98.95; FL: Acc=96.88\%, F1=97.92 \\

    \cite{n16} & UNSW-NB15, AWID, NSL-KDD (combined; with 30\% poisoning attack simulation) & Network traffic features and packet-level metadata & Asynchronous Advantage Actor-Critic (A3C) deep reinforcement learning (actor-critic) & N/A & N/A & Accuracy: 98.68\%, Precision: 98.4\%, Recall: 98.9\%, FP rate: 1.59\% \\

    Pro-posed & Bot-IoT (25\% \& 30\% subset) & 34 features of network traffic & A2C, Q-Learning \& Meta RL & Post-hoc weight modification & Teacher: A2C \newline Student: A2C & Best: \textbf{A2C (25\% subset)} Accuracy=99.60\%, F1=99.80\% \\

    \hline \hline
    \end{tabular}
    \label{lit_summary}
\end{table*}

\section{Methodology}\label{methodology}
\begin{figure*}[htbp]
  \centering
  \includegraphics[width=1.85\columnwidth]{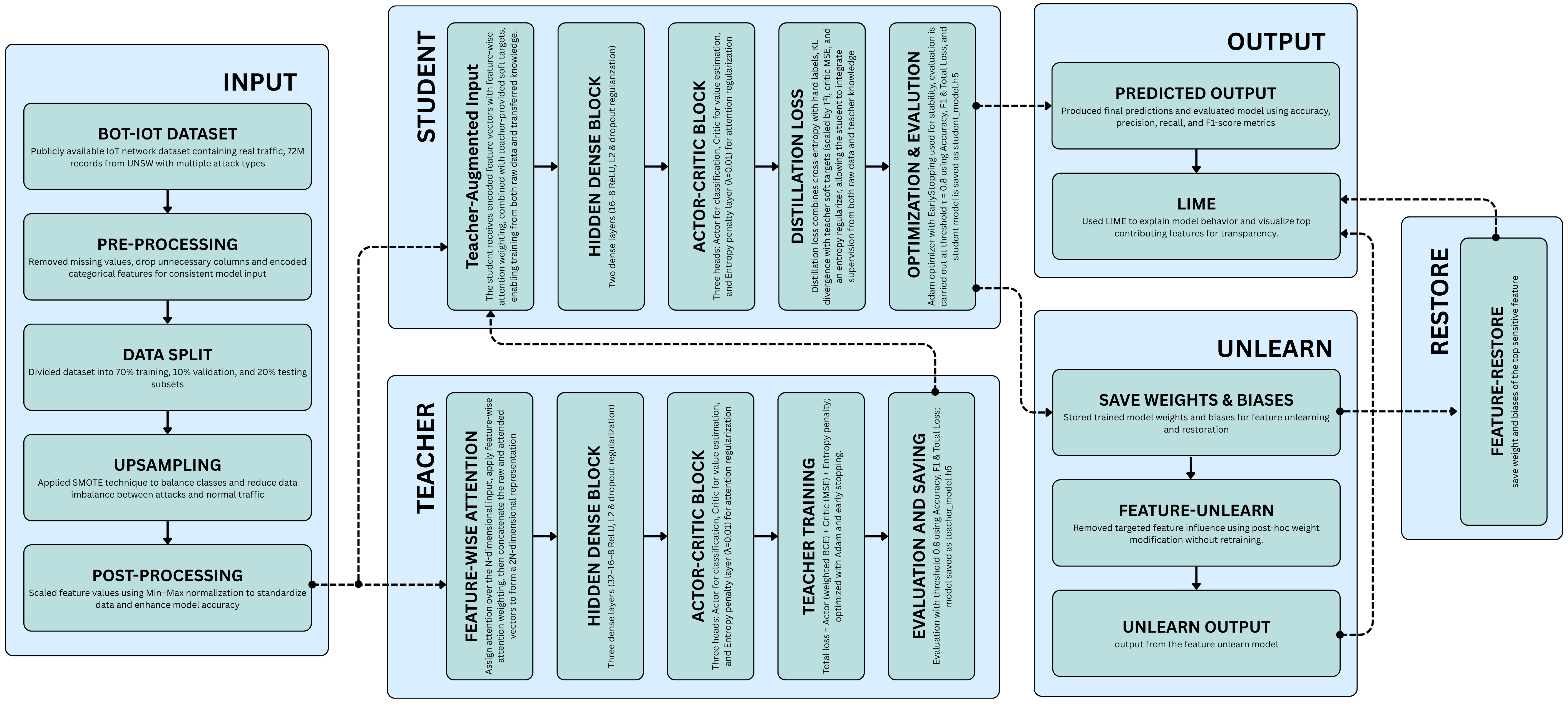}
  \caption{A detailed architecture of the DiRLU framework, where the teacher transfers knowledge to the student via knowledge distillation, integrated with post-hoc feature unlearning module for data privacy also reintroduce unlearned features for scalability}
  \label{fig1}
\end{figure*}

\begin{figure*}[htbp]
  \centering
  \includegraphics[width=1.85\columnwidth]{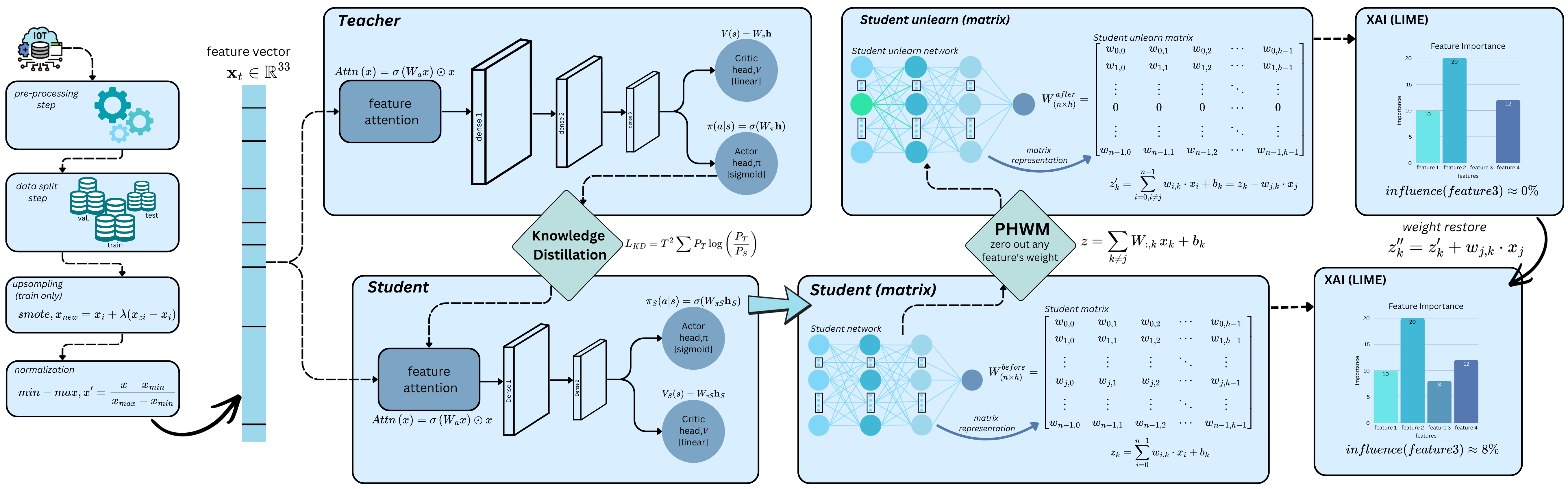}
  \caption{Schematic overview of the proposed DiRLU framework based on knowledge distillation and post-hoc weight modification (PHWM). A teacher actor–critic network transfers knowledge to a compact student model, after which PHWM selectively zeros the weights associated with input features in the student’s weight}
  \label{schematicfig}
\end{figure*}

This research aims to identify attack traffic accurately, even when attack traffic dominates benign traffic, by utilizing a large-scale dataset. Besides, we seek to develop a compact model that performs efficiently without requiring extensive computational resources. Additionally, the study focuses on the ability to selectively forget sensitive data whenever needed, ensuring enhanced data security and privacy. Figure-\ref{fig1} illustrates the detailed architecture of the proposed model, DiRLU. The process begins with IoT devices generating network traffic data. This data then goes through preprocessing and feature extraction using engineering techniques. The processed data was split into three portions for training, testing, and validation. In the modeling stage, we use reinforcement learning methods such as A2C, Q-learning and Meta-RL. A feature unlearning module works with these RL models to selectively remove sensitive or outdated features from the system. To make the trained model lighter and faster, we apply knowledge distillation. The distilled model then produces the output (attack or benign). Using XAI (LIME), we explain the model’s decisions so they are transparent and trustworthy. Finally, performance metrics are used to evaluate the effectiveness of the models and to measure the effect of the model. This entire workflow demonstrates a dynamic approach to IoT security. However, figure-\ref{schematicfig} presents the schematic overview of the proposed DiRLU model. It illustrates the unlearning mechanism as a direct matrix-level intervention $(W_{:,j}=0)$, where the weight rows corresponding to selected features are set to zero to remove their influence without retraining. The figure also illustrates the knowledge distillation process, showing how the student model learns from the teacher's softened predictions. It also highlights how unlearning updates feature contributions while making decisions.

The following sections provide a detailed explanation of the various phases of the proposed framework. Since multiple symbols are used in later sections, a list of commonly used symbols is included in table-\ref{symbol_table} to enhance the clarity of this paper.

\begin{table}[!ht]
    \centering
    \caption{Mathematical Symbols for Proposed Architecture}
    \begin{tabular}{>{\raggedright\arraybackslash}p{1cm} p{7cm}}
        \toprule
        \textbf{Symbol} & \textbf{Description} \\
        \midrule
        \multicolumn{2}{l}{\textbf{Normalization}} \\
        \midrule
        $x$ & Original value of a feature\\
        $x_{\text{min}}$ & Min value of a feature in the dataset \\
        $x_{\text{max}}$ & Max value of a feature in the dataset \\
        $x'$ & Normalized value after Min-Max scaling \\
        \midrule
        \multicolumn{2}{l}{\textbf{Actor Network}} \\
        \midrule
        $\pi_\theta(a|s)$ & Policy function of the actor network with parameters $\theta$ \\
        $\theta$ & Parameters of the actor network \\
        $L_{\text{actor}}$ & Loss function for the actor network \\
        $A_t$ & Advantage function at time step $t$ \\
        $\alpha$ & Learning rate for the actor network \\
        $\nabla_\theta$ & Gradient with respect to actor parameters \\
        \midrule
        \multicolumn{2}{l}{\textbf{Critic Network}} \\
        \midrule
        $V_\phi(s_t)$ & Value function of the critic network for state $s_t$ \\
        $L_{\text{critic}}$ & Loss function for the critic network \\
        $R_t$ & Expected cumulative reward at time $t$ \\
        $r_t$ & Immediate reward at time step $t$ \\
        $\gamma$ & Discount factor in reinforcement learning \\
        $\beta$ & Learning rate for the critic network \\
        \midrule
        \multicolumn{2}{l}{\textbf{Knowledge Distillation}} \\
        \midrule
        $P_T(i)$ & Soft target probability from the teacher model for class $i$ \\
        $z_i$ & Logit (output) from the teacher model for class $i$ \\
        $T$ & Temperature parameter for softmax smoothing \\
        $P_S(i)$ & Student model’s predicted probability for class $i$ \\
        $y_i$ & Ground truth label for class $i$ \\
        $L_{\text{CE}}$ & Cross-entropy loss for true labels \\
        $L_{\text{KD}}$ & Distillation loss (KL divergence with soft targets) \\
        $L$ & Total loss function\\
        \midrule
        \multicolumn{2}{l}{\textbf{Evaluation}} \\
        \midrule
        $y_i$ & Ground truth label for class $i$ \\
        $\hat{y}_i$ & Predicted probability for class $i$ by the model  \\
        $n$ & The number of samples\\ 
        \bottomrule
    \end{tabular}
    \label{symbol_table}
\end{table}

\subsection{Data Description}
The BoT-IoT publicly available dataset was built in a real network environment in the Cyber Range Lab of the University of New South Wales (UNSW) Canberra \cite{data_des_1}. The environment was set in such a way that contained both normal and botnet network traffic. In the dataset, normal traffic is dominated by attack traffic. The dataset is heavily imbalanced, which reflects real-world cyberattack scenarios. They captured traffic using Wireshark, and the total size of pcap files is 69.3 GB, with more than 72 million records and 35 features. The Bot-IoT dataset includes different attacks such as DDoS, DoS, OS and Service Scan, Keylogging and Data exfiltration attacks, which are explained in Table-\ref{table_attackName}.

\begin{table*}[!ht]
    \caption{Different cyber-attacks listed in the Bot-IoT dataset}
    \centering
    \begin{tabular}{p{4.5cm} p{10cm}}
    \hline \hline
    \textbf{Attack Name} & \textbf{Description}\\ \hline
     
     Denial of Service (DoS) & An attempt to make a machine or network resource unavailable to its intended users\\ 

     Distributed Denial of Service (DDoS) & Multiple compromised systems attack a target, causing denial of service for users\\

     Reconnaissance & Activities aimed at gathering information about a network for planning future attacks\\

     Information Theft & Unauthorized access and retrieval of sensitive information\\

     Data Exfiltration & Unauthorized transfer of data from a computer or other device\\

     OS and Service Scan & This technique reveals systems and open service (HTTP, FTP, or SMTP) vulnerabilities\\

     Keylogging & This is a secret recording of a user’s keystrokes to capture sensitive information\\
     
     \hline \hline
    \end{tabular}
    \label{table_attackName}
\end{table*}

In Table-\ref{table_data_des}, we listed all populated features (total 31) along with their description.

\begin{table}[!ht]
    \centering
    \caption{Feature description of the Bot-IoT dataset}
    \begin{tabular}{l l}
    \hline
        \textbf{Feature Name} & \textbf{Feature Description} \\ \hline
        pkSeqID & Packet number \\
        stime & Packet capture start time \\ 
        flgs & Flow state flags seen in transactions \\
        flgs\_number & Numeric num. of flgs\\
        proto & Protocol used in the flow\\
        saddr & Source IP\\
        sport & Source port\\
        daddr & Destination IP\\
        dport & Destination port\\
        pkts & Total packet count\\
        bytes & Total bytes in the flow\\
        state & Packet state\\
        state\_number & Numeric num. of state\\
        ltime & Packet capture last time \\ 
        seq & Packet sequence number\\
        dur & Total duration\\
        min & Min duration of aggregated records\\
        max & Max duration of aggregated records\\
        mean & Average duration of total time\\
        stddev & Standard deviation of all records\\
        sum & Total duration of all records\\
        spkts & Source to destination packet count\\
        dpkts & destination to Source packet count\\
        sbytes & Byte count of source to destination packets\\
        dbytes & Byte count of destination to Source packets\\
        rate & Packets per seconds in the flow\\
        srate & Packets per seconds from source to destination\\
        drate & Packets per seconds from destination to Source\\
        attack & Label: 0 for Normal and 1 for Attack\\
        category & Network traffic category\\
        subcategory & Network traffic sub-category\\ \hline
    \end{tabular}
    \label{table_data_des}
\end{table}

\subsection{Data Pre-Processing}
Data pre-processing is required to ensure that the data fit properly into the proposed model as it standardizes, normalizes, and handles missing values. This alignment optimizes the model performance and accuracy by furnishing data with uniform, high-quality, and ideal inputs.

Firstly, we have dropped the \textit{category and subcategory} column as we do not require them for classification. So, out of 35 features, we worked with 33 features; among them, we selected 32 features for further preprocessing. However, the column named \textit{attack} is used for classification.

\subsubsection{Feature Pruning}
To have a refined and clean dataset for training the model, we removed all missing or nan values. Some of the features didn't even have any values except nan. So, after removing all nan values, we got 27 columns and 15,000,000 rows. The completely empty features are \textit{smac, dmac, soui, doui, sco, }and\textit{ dco}.

\subsubsection{Label Encoding}
Label encoding is an essential technique for converting category input to numerical values, which makes it suitable for deep learning algorithms. For example,  we transformed the "attack" category, assigning the value 1 for attack and 0 for not attack. We utilized the \textit{LabelEncoder} function from the \textit{sklearn} library. This conversion allows the model to process the categorical data and use them for classification purposes.

\subsection{Data Split}
Splitting the dataset into a train, validation, and test subset is crucial for training, validating, and testing the model. We have taken 25\% of the total dataset, whereas previous state-of-the-art research has taken only 5\% \cite{n3}. We got 15,000,000 (Fifteen million) data points in 25\% of the whole dataset. We use a random splitting technique to split the dataset into three parts: training, validating, and testing. We took 70\% of data points for training, 10\% for validating \& tuning hyper-parameters, and another 20\% for testing the model. We randomly choose the 70:20:10 (train:test:validation) ratio for splitting the dataset. However, after splitting, we got 10,500,000 (Ten million five hundred thousand) data points for training the model, 1,500,000 (One million five hundred thousand) data points for tuning hyper-parameters \& validating the model, and 3,000,000 (Three million) data points for testing the model's performance.

\subsection{Data Augmentation}
We have taken 25\% data points from the BOT-IOT dataset, which gives us a total of 15,000,000 rows. Among those rows, we found attack rows 14,992,430 (Fourteen million, nine hundred ninety-two thousand, four hundred thirty) and the remaining 7,570 (Seven thousand five hundred seventy) non-attack rows. The dataset we have chosen for this research is highly imbalanced, and the ratio of attack vs non-attack is 98:2. So, to balance the dataset, we have used a technique called SMOTE.

SMOTE (Synthetic Minority Oversampling Technique) is an oversampling preprocessing technique used to address class imbalances in the dataset \cite{brandt2021comparative}. It generates synthetic data points by selecting a minority class and creating new data points along the line segments, attaching them to their closest neighbors. This method introduces diversity in the artificial samples, helping to create a non-biased dataset. SMOTE has been widely adopted in machine learning applications to enhance classification performance in imbalanced datasets \cite{chawla2002smote}.

\begin{figure}[htbp]
  \centering
  \includegraphics[width=0.95\columnwidth]{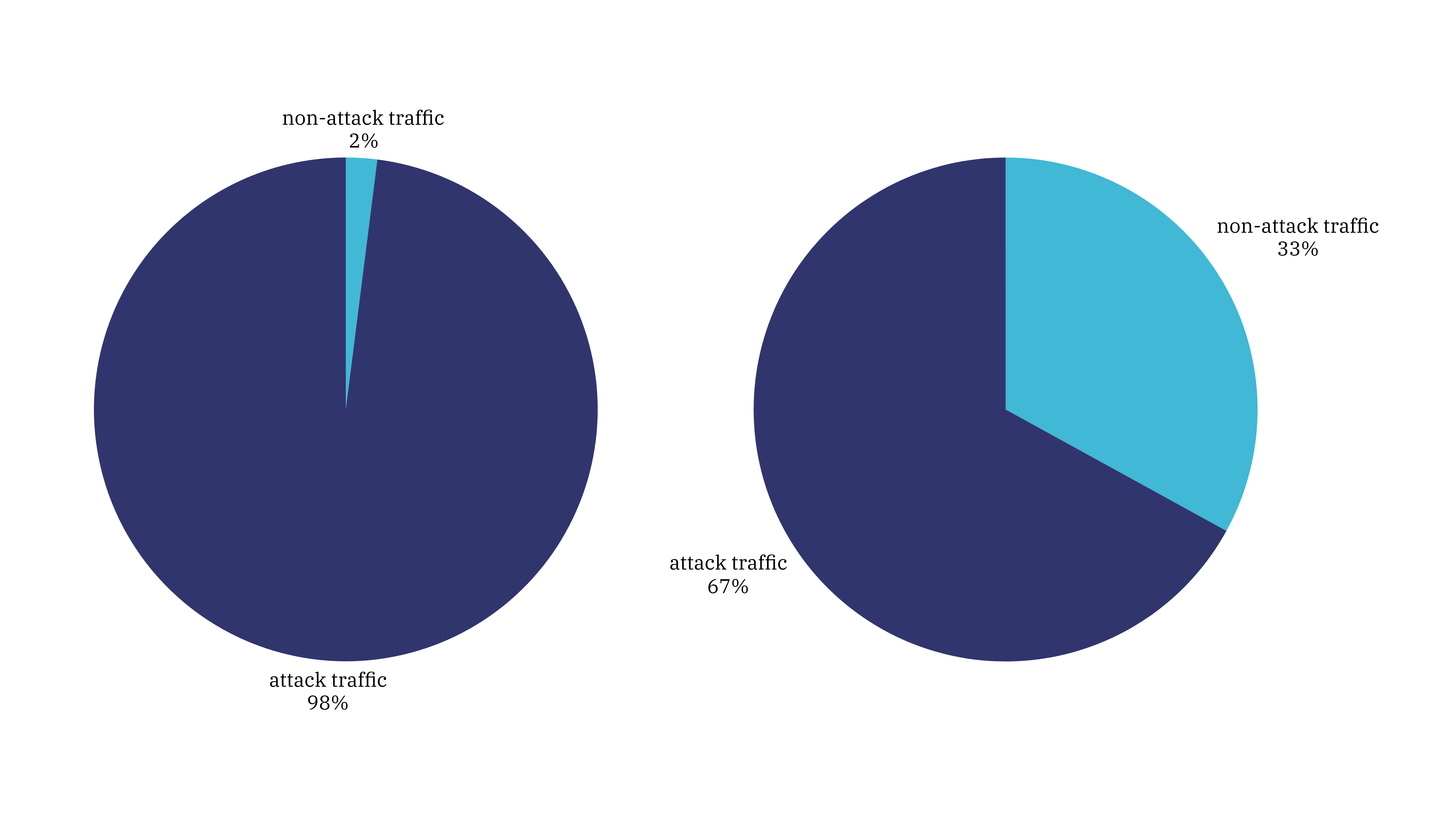}
  \caption{Comparison of before and after applying SMOTE}
  \label{fig2}
\end{figure}

After applying smote to the imbalance training dataset, we got 7,035,000 (Seven million thirty-five thousand) rows for the attack class (67\%) and another 3,465,000 (Three million four hundred sixty-five thousand) rows for the non-attack class (33\%), which makes the dataset balanced and non-biased. In Figure-\ref{fig2}, we have represented the ratio and effect of before \& after applying smote.

\subsection{Data Normalization}
Data normalization is a vital pre-processing technique required for scaling feature values. In the dataset, some feature values are large, and some are very small. Basically, it reshapes numerical values to fit into standard scale values. In this research, we applied deep learning techniques, so we used the min-max normalization technique to have values between 0 and 1. Recent studies have proved that min-max normalization can significantly improve the accuracy of deep learning models across various applications \cite{deepa2022epileptic}. The Min-Max function normalized the data using equation-\ref{eq:1}.

\begin{equation}\label{eq:1}
x' = \frac{x - x_{\text{min}}}{x_{\text{max}} - x_{\text{min}}}
\end{equation}

In equation-\ref{eq:1}, \(x\) is the original value, \(x_{\text{min}}\) is the minimum \& \(x_{\text{max}}\) is the maximum value in the dataset, and \(x'\) is the normalized value.

\subsection{Reinforcement Learning}
Reinforcement learning is a type of machine learning where the model learns by interacting with its environment and receiving rewards or penalties for each of its actions, similar to trial and error \cite{RF_IBM}. Unlike traditional deep learning, it does not rely on labeled data but improves through feedback. It is beneficial for solving complex problems like IoT Security \cite{uprety2020reinforcement}.

\begin{figure}[htbp]
  \centering
  \includegraphics[width=0.8\columnwidth]{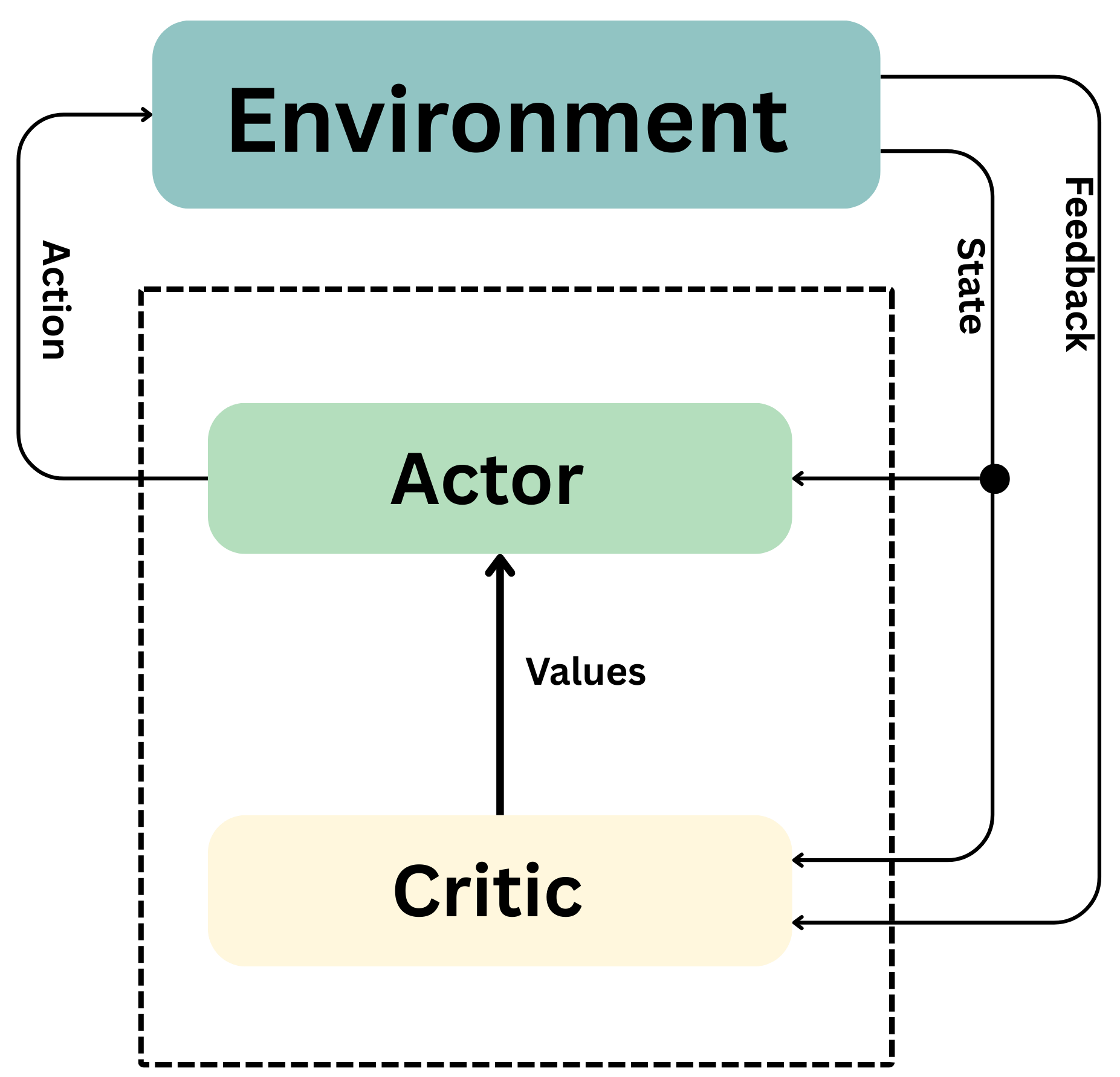}
  \caption{Actor-Critic model architecture}
  \label{RF_arc}
\end{figure}

We have used the actor-critic model because of its enhanced stability compared to traditional reinforcement learning approaches by combining policy optimization with value estimation. In the Fig-\ref{RF_arc}, we represented a simple architecture of the actor-critic model. The actor–critic framework relies on two connected neural networks that work together to improve decision-making. The actor network acts as the policy generator, selecting actions based on the current state, while the critic network evaluates those actions by estimating their quality and long-term value. During training, the actor updates its policy using feedback from the critic, which in turn refines its value estimates based on observed rewards. This interaction creates a balanced learning setup that encourages exploration of new strategies. A key element in this process is the temporal difference error, which measures the gap between predicted and actual rewards and serves as the main learning signal for both networks \cite{actor_critic_tensorflow}. The actor learns by adjusting its policy through gradient ascent on expected rewards, while the critic reduces prediction errors using value function approximation. Through repeated updates, the system gradually converges to a stable policy that maximizes cumulative rewards and maintains reliable learning dynamics.

From equations \ref{policy_function}-\ref{critic_update}, we have described the mathematical formulas of the actor-critic model, including both forward propagation and backpropagation update rules.

\begin{equation}\label{policy_function}
\pi_\theta(a|s) = P(a|s; \theta)
\end{equation}
In equation-\ref{policy_function} is Actor Policy Function is shown where the probability of taking action $a$ in state $s$, and $\theta$ denotes the actor network parameters.

\begin{equation}\label{actor_loss}
L_{actor} = -\mathbb{E}[\log \pi_\theta(a_t|s_t) \cdot A_t]
\end{equation}
In equation-\ref{actor_loss} shows the actor loss function, where $A_t$ is the advantage function and $\mathbb{E}[\cdot]$ represents the expected value operator.

\begin{equation}\label{value_function}
V_\phi(s_t) = \mathbb{E}[R_t|s_t]
\end{equation}
In equation-\ref{value_function} shows Critic Value Function where critic parameters is $\phi$, and $R_t$ denotes the expected cumulative reward.

\begin{equation}\label{critic_loss}
L_{critic} = \mathbb{E}[(r_t + \gamma V_\phi(s_{t+1}) - V_\phi(s_t))^2]
\end{equation}
In equation-\ref{critic_loss}, the critic loss function is shown, where $V_\phi(s_t)$ is the state value function, $\gamma$ represents the discount factor, and $r_t$ is the immediate reward at time step $t$.

\begin{equation}\label{actor_update}
\theta \leftarrow \theta + \alpha \nabla_\theta \log \pi_\theta(a_t|s_t) \cdot A_t
\end{equation}
In equation-\ref{actor_update} presents the actor parameter update rule, where $\alpha$ is the learning rate and $\nabla_\theta$ represents the gradient with respect to actor parameters.

\begin{equation}\label{critic_update}
\phi \leftarrow \phi - \beta \nabla_\phi L_{critic}
\end{equation}
In equation-\ref{critic_update}, the critic parameter update is shown, where $\beta$ is the critic learning rate and $\nabla_\phi$ denotes the gradient with respect to critic parameters.

We used actor-critic reinforcement learning in a teacher-student framework for knowledge distillation. In this design, the teacher model has two outputs: the actor makes binary classification decisions using a sigmoid function, and the critic provides continuous value estimates with a linear function. We added an attention mechanism that learns to focus on all the input features with an entropy regularization. We also implemented shared hidden layers, L2 and dropout regularization for better generalization. We use weighted loss functions to handle class imbalance so that underrepresented classes get fair attention during training. Overall, the model jointly optimizes three goals: accurate classification by the actor, precise value estimation by the critic, and balanced attention. This multi-objective setup helps the teacher learn rich representations, which are then passed on to the student model for better learning.

\subsection{Knowledge Distillation}
\begin{figure}[htbp]
  \centering
  \includegraphics[width=1.0\columnwidth]{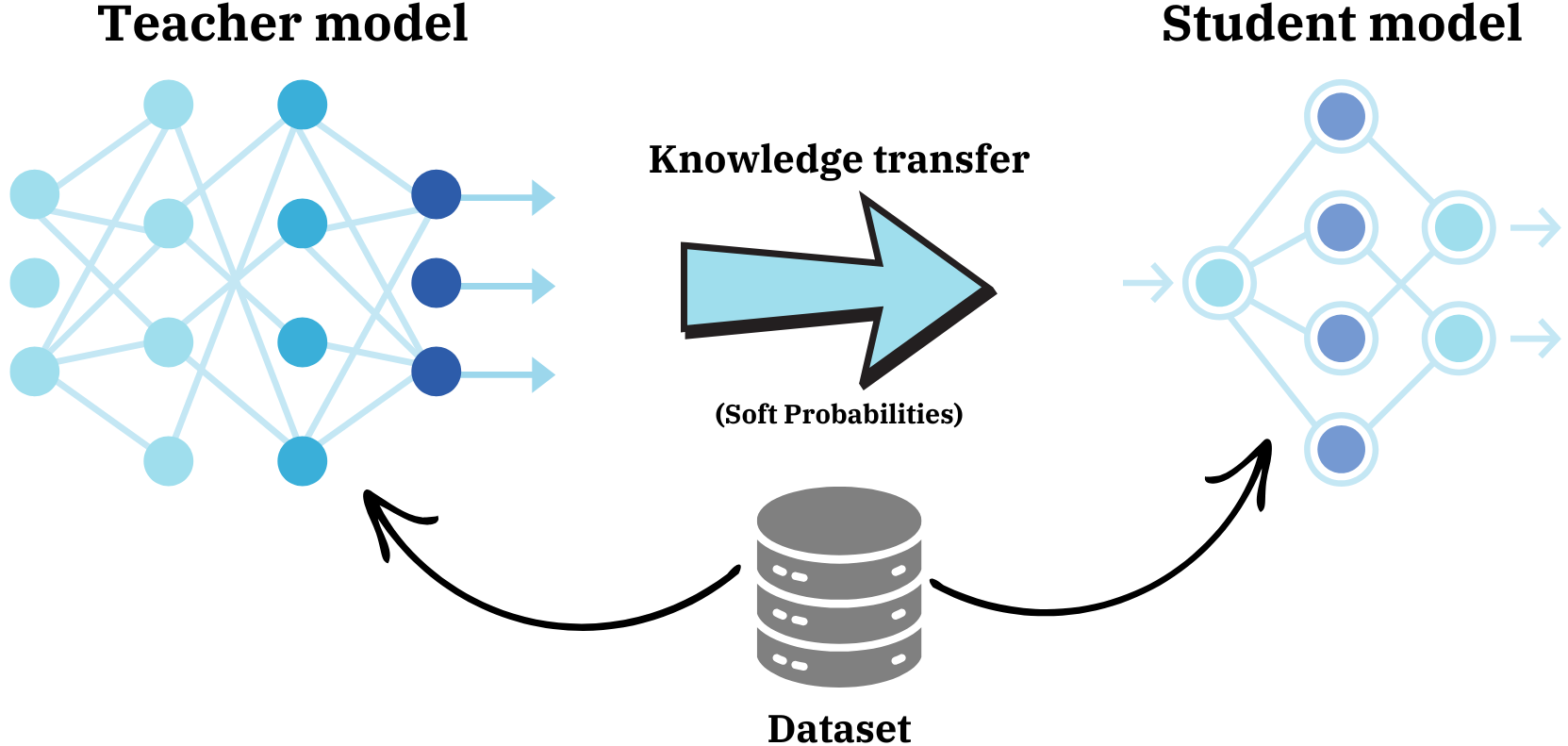}
  \caption{Architecture of knowledge distillation (KD) framework}
  \label{fig4}
\end{figure}

Knowledge Distillation (KD) is a model compression technique where a large, complex teacher model transfers its knowledge to a smaller, more efficient student model without significant loss in performance. In this method, the teacher model is initially trained on a dataset, and its output, such as soft probabilities (logits) or intermediate feature representations, is used to assist the student model's training. The student model is optimized using a distillation loss function, which combines standard loss (cross-entropy) with a term encouraging the student to mimic the teacher's outputs. The benefit of the KD is that it reduces the size of networks while maintaining high accuracy and faster inference, making it appropriate for deployment on resource-constrained devices such as mobile phones and edge computing platforms. This technique has been widely adopted in security applications, improving the robustness of AI models against adversarial attacks while maintaining efficiency \cite{gou2021knowledge}. We have used KD to classify malware and begin traffic with low resources. In Fig-\ref{fig4}, we have included an architecture diagram of the knowledge distillation model.

\subsubsection{Teacher Model}\label{Teacher Model}
We implemented a teacher model within a Knowledge Distillation (KD) framework, where a larger Actor-Critic model transfers its knowledge to a smaller one (student model). The teacher model learns to map input data to output probabilities, which are then used to transfer knowledge to a smaller one. The teacher model produces soft targets $P_T$ by applying a temperature-scaled softmax function to the logits:

\begin{equation}\label{eq:2}
P_T(i) = \frac{\exp(z_i / T)}{\sum_{j} \exp(z_j / T)}
\end{equation}

In equation-\ref{eq:2}, $z_i$ is the logit (output) from the teacher model and 
$T$ is the temperature hyperparameter, which controls the smoothness of the probability distribution. Higher $T$ produces smoother probability distributions.

The teacher model is an attention-enhanced actor-critic architecture. It includes a feature-level attention mechanism that learns from all the individual input features. The attended representation is concatenated with the original input and passed through three shared dense layers with ReLU activation and L2 regularization. The actor's head generates probability scores using a sigmoid activation, while the critic's head estimates value functions using a linear activation. An additional entropy penalty regularizes the attention weights to prevent overly confident focus on a few features. Class weights are dynamically computed and incorporated into the custom binary cross-entropy loss function to address class imbalance. Furthermore, the model uses early stopping on validation accuracy to avoid overfitting and restore the best-performing weights. Training continues until early stopping is triggered, typically within 25–30 epochs. Table-\ref{table_4} provides a complete list of the hyperparameters and their corresponding values used to improve the teacher model's performance.

\begin{table}[!ht]
    \centering
    \caption{Hyperparameters used in teacher model}
    \begin{tabularx}{0.9\columnwidth}{l X} 
    \hline
        \textbf{Hyperparameter} & \textbf{Value} \\ 
            \hline
            Model & Actor-Critic \\
            Number of Hidden Layers & 3 \\
            Activation Function (Hidden Layers) & ReLU \\
            Activation Function (Actor Output) & Sigmoid \\
            Activation Function (Critic Output) & Linear \\
            Regularization & L2 \& Entropy Penalty\\
            Actor Loss Function & Weighted BSE\\
            Critic Loss Function & MSE\\
            Optimizer & Adam \\
            Epoch Number & Early stopping (max 30)\\
            Batch Size & 512 \\
            Evaluation Metric & Accuracy \& F1 Score \\
            \hline
    \end{tabularx}
    \label{table_4}
\end{table}

To make predictions, probability scores from the actor's head are binarized using a 0.8 threshold, and the F1 score is calculated to evaluate classification performance. This trained teacher model is the foundation for knowledge transfer, guiding the development of a smaller, more efficient student model with similar decision boundaries and attention-driven behaviour.

\subsubsection{Student Model}\label{Student Model}
A student model is a smaller, more efficient Actor-Critic model trained to replicate the behaviour of a larger and more complex teacher model. It learns using the teacher's softened predictions, allowing it to perform well while being lightweight and faster. With appropriate distillation techniques, student models can accurately mimic the behaviour of the teacher model while drastically reducing the model size and computational complexity, making them ideal for real-time applications in mobile and edge computing \cite{chen2022improved}. The student model mimics the teacher's behavior by optimizing a total loss that combines standard cross-entropy (for true labels) and distillation loss (KL divergence with the teacher's soft targets).\\

The Cross-Entropy Loss for true labels $(L_{\text{CE}})$,

\begin{equation}\label{eq:3}
L_{\text{CE}} = -\sum_{i} y_i \log P_S(i)
\end{equation}

In equation-\ref{eq:3}, $ y_i$ is the ground truth label and $P_S(i)$ is the student model's predicted probability.\\

Distillation Loss (KL divergence with soft targets) $(L_{\text{KD}})$,

\begin{equation}\label{eq:4}
L_{\text{KD}} = T^2 \sum_{i} P_T(i) \log \frac{P_T(i)}{P_S(i)}
\end{equation}

In equation-\ref{eq:4}, $T$ is temperature parameter which controls the smoothness of the probability distribution, $P_T(i)$ is probability output from the teacher model of class $i$ and $P_S(i)$ is probability output from the student model of class $i$.\\

The total loss function $(L)$,

\begin{equation}\label{eq:5}
L = \alpha L_{\text{CE}} + (1 - \alpha) L_{\text{KD}}
\end{equation}

In equation-\ref{eq:5}, $\alpha$ is a weighting factor that balances the cross-entropy loss $L_{\text{CE}}$ and the distillation loss $L_{\text{KD}}$.\\

In this research, the student model consists of a smaller actor-critic model. This student model is smaller and faster than the teacher model. The actor helps predicts the final class (benign or attack), and the critic, which gives a value or score to help guide learning.

We use an attention mechanism to make the model focus on all the input features. These attention-weighted inputs are combined with the original input and passed through two hidden layers. Each layer uses the ReLU activation function, and we apply L2 regularization to reduce overfitting. The final part of the model has two outputs: the actor output uses a sigmoid function for classification, and the critic output uses a linear function to estimate value. We also apply an entropy penalty on the attention weights to ensure the model doesn’t focus too much on only a few features. We train the model using the Adam optimizer with a batch size 32. The actor output is trained with a custom loss function that combines two types of loss: binary cross-entropy, which checks how well the predictions match the actual labels, and KL divergence, which helps the student learn from the soft outputs of the teacher model. We use a temperature value of 2.0 to smooth the teacher’s predictions and an alpha value 0.5 to balance the two losses. For the critic output, we use mean squared error with class weights to handle imbalance in the data. We also include the entropy penalty in the total loss to balance the attention distribution. To prevent overfitting, we apply early stopping on validation accuracy, stopping training once the model no longer shows improvement. The hyperparameters and their corresponding values used to enhance the student model’s performance are summarized in Table-\ref{table_5}.

\begin{table}[!ht]
    \centering
    \caption{Hyperparameters used in student model}
    \begin{tabularx}{0.9\columnwidth}{l X} 
    \hline
        \textbf{Hyperparameter} & \textbf{Value} \\ 
        \hline
        Model & Actor-Critic \\
        Number of Hidden Layers & 2 \\
        Activation Function (Hidden Layers) & ReLU \\
        Activation Function (Actor Output) & Sigmoid \\
        Activation Function (Critic Output) & Linear \\
        Regularization & L2 \& Entropy Penalty\\
        Actor Loss Function & Distillation Loss (CE+KL)\\
        Critic Loss Function & Weighted MSE\\
        Temperature & 2.0 \\
        Alpha & 0.5 \\
        Optimizer & Adam \\
        Epoch Number & Early stopping (max 30)\\
        Batch Size & 512 \\
        Evaluation Metric & Accuracy \& F1 Score \\
        \hline
    \end{tabularx}
    \label{table_5}
\end{table}


\subsection{Un-learning}
Feature unlearning is a technique designed to remove the influence of specific features from a trained model, ensuring that once removed, those features no longer affect the model’s outputs.

The Price of Forgetting \cite{PriceOfForgetting} introduced an economic approach by offering payments to users for keeping their data, helping servers reduce costs while still respecting user privacy. Also, unlearning requires retraining the model from scratch, which is computationally expensive \cite{LearntoForget}. While our approach plays an important role in meeting privacy regulations, mitigating security risks (such as eliminating data), and enabling efficient model updates without the need for full retraining. Key benefits include regulatory compliance by securely removing sensitive data, cost and time savings from avoiding comprehensive retraining, and greater transparency and trust in AI systems. Recent developments in feature unlearning have introduced methods such as data pruning and selective retraining, which allow models to effectively erase the impact of targeted features while preserving overall performance \cite{zhang2023machine}. In practice, this balance of privacy, efficiency, and model reliability is critical for real-world applications.

\begin{figure}[htbp]
  \centering
  \includegraphics[width=0.92\columnwidth]{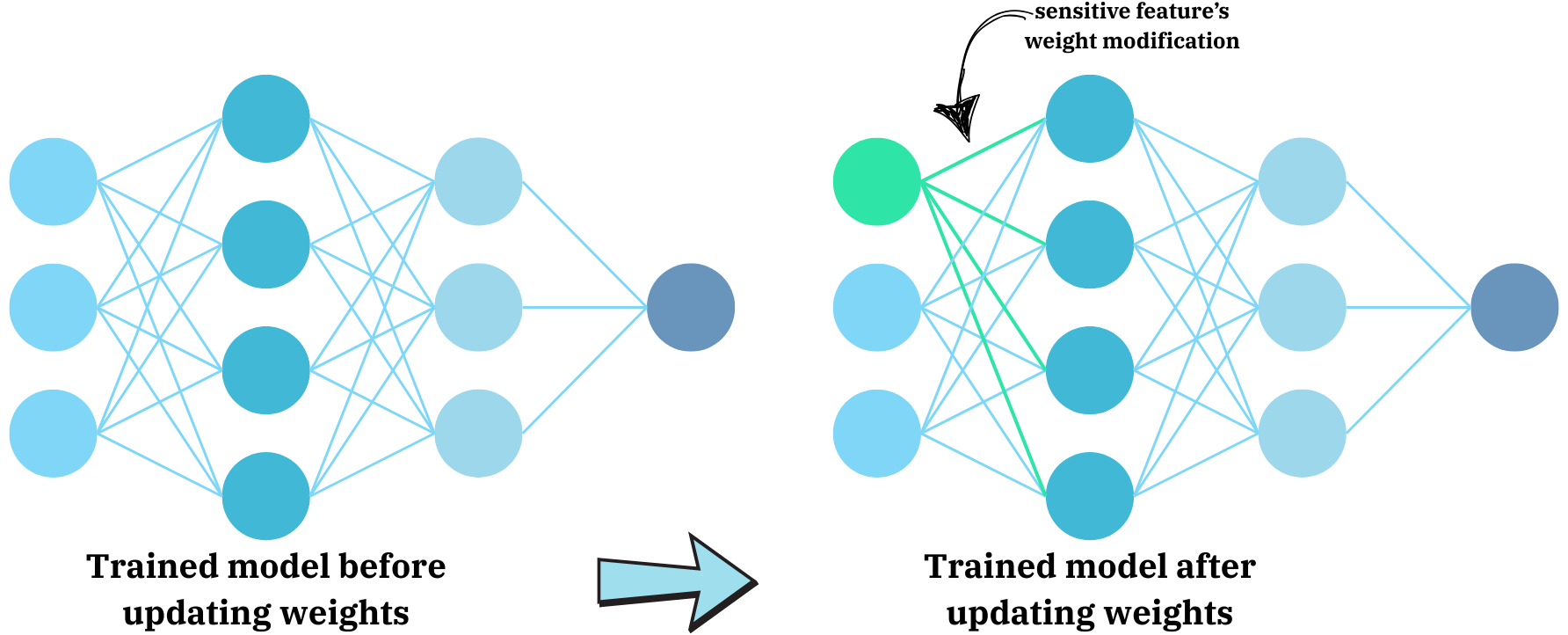}
  \caption{Architecture of feature unlearning process}
  \label{PHWM}
\end{figure}

One practical method for feature unlearning is post hoc weight modification (PHWM), which enables models to “forget” sensitive data or features without retraining from scratch. Instead of rebuilding the entire network, PHWM simply sets the weight of a given feature to zero in the first layer, effectively removing its influence. By adjusting a single weight, the model can quickly forget targeted information, avoiding the high cost and time required for full retraining \cite{hakemi2025post}. Figure-\ref{PHWM} illustrates this process, showing how the method streamlines unlearning while maintaining the model’s efficiency.

We applied PHWM to unlearn a feature $x_j$ by removing its contribution in the trained network. Consider the first dense layer,
\begin{equation}
z = Wx+b,
\end{equation}
where $x$ denotes the input feature vector, $W$ and $b$ are the weight matrix and bias vector of the first layer, and $z$ is the corresponding output. To forget the feature $x_j$, we set all first-layer connections associated with that feature to zero:
\begin{equation}
W_{:,j} = 0.
\end{equation}
With this modification, the first-layer’s output becomes
\begin{equation}
z = \sum_{k \neq j} W_{:,k}\,x_k+b,
\end{equation}
which no longer contains any term involving $x_j$. As a result, changing $x_j$ while keeping the remaining features fixed does not change $z$, meaning the first-layer representation is independent of the unlearned feature.

So, after this update, the removed feature cannot affect the first-layer activations. Because every later layer uses only $z$ (or values computed from $z$), the feature $x_j$ remains ignored throughout the network. In other words, once the first-layer connections are set to zero, the model’s later computations depend only on the remaining features.

\subsubsection{Teacher Unlearn}
The teacher unlearning process involves selectively removing the second most influential feature from the dataset, as determined by its impact on the teacher model. Using XAI, we identified \textit{flgs} as this feature and removed it from the input data to implement unlearning. In addition, we set the corresponding weights in the model’s first layer to zero, ensuring the feature no longer influenced predictions. The modified weight matrix was then reloaded into the model to apply these changes. To evaluate the effect, we measured performance using accuracy and F1-score. This approach improves interpretability by removing redundant or sensitive features while keeping computations efficient. Enforcing unlearning at both the data and model levels ensures the model no longer relies on the removed feature.

\subsubsection{Student Unlearn}
The student unlearning process involves removing the second most influential feature from the dataset that impacts the student model. Using XAI, we identified \textit{flgs} as this feature. It was excluded from the input data to ensure the model no longer learned from or relied on it for predictions. In addition, the corresponding weights in the first layer were set to zero, preventing any influence from the feature. The updated weight matrix was then reloaded into the student model to confirm the feature’s removal. To assess the effect of these changes, we re-evaluated the model using accuracy and F1-score. This approach enhances interpretability by eliminating unnecessary or sensitive features while preserving computational efficiency.

\subsection{Explainable Artificial Intelligence (XAI)}
Explainable AI (XAI) in machine learning and deep learning focuses on making model decisions more transparent and easier to understand. It demonstrates how models arrive at their predictions, which helps build trust and ensures fairness. XAI is also valuable for assessing feature importance by showing how different factors influence model outcomes, an essential step for tuning models and improving performance. By improving interpretability, XAI supports accountability and strengthens the reliability of AI systems. Common approaches include local and global interpretability methods such as SHAP and LIME, which provide insights into both individual predictions and overall model behavior, making it easier to detect biases or errors in decision-making \cite{ribeiro2016should}. Recent work has also applied XAI to anomaly detection in NFV systems, where masked autoencoders are combined with explainability techniques to enhance interpretability and support effective fault localization \cite{AnomalyNFV}.

In our work, we applied LIME (Local Interpretable Model-agnostic Explanations) to gain a more in-depth understanding of how the model makes its projections. LIME works by generating minor variations around a given data point and analyzing how the model reacts, offering insights into the most influential features. It has been widely used across machine learning tasks to improve transparency and help identify potential biases or errors \cite{gilpin2018explaining}. For our experiment, we randomly selected a single test instance and used LIME to rank the top 12 features influencing the model’s prediction. The explanation mode was set to classification, and discretization was disabled to retain continuous feature values. The Results were visualized interactively using show\_in\_notebook(). Using LIME, we gained valuable insights into which features played a key role in the model's classification.

\subsection{Proposed Algorithm}
In algorithm-\ref{pseudocode}, we explain the pseudocode for the proposed model's architecture in detail. We built the model using Python (v3.12) and different libraries, such as tensorflow, pandas, lime, matplotlib, etc., on Visual Studio Code IDE.

\begin{algorithm}[!htb]
\caption{Pseudocode of DiRLU}
\resizebox{0.96\linewidth}{!}{%
\begin{minipage}{\linewidth}
\small
\begin{algorithmic}[1]

\Function{Preprocess}{df}
  \State $df_c \gets \textproc{Clean}(df)$
  \State $df_b \gets \textproc{ApplySmote}(df_c)$
  \State $df_{tr}, df_{val}, df_{te} \gets \textproc{Split}(df_b)$
  \State \Return $df_{tr}, df_{val}, df_{te}$
\EndFunction

\vspace{0.4em}

\Function{TrainA2CTeacher}{df\_train}
  \State $teacher \gets \textproc{A2C()}$
  \State $cls\_weights \gets \textproc{CompClsWeights}(df\_train)$
  \While{NOT \textproc{Converged}(teacher, EarlyStop)}
    \State $(x, y) \gets \textproc{Sample}(df\_train)$
    \State $L \gets \textproc{WeightedBCE}(actor(x), y, cls\_weights)$ 
    \Statex \hspace{1.5cm}$+ \textproc{MSE}(critic(x), y)$ 
    \Statex \hspace{1.5cm}$+ \lambda_{attn}\cdot \textproc{Entropy}(attention)$
    \State \textproc{UpdateA2C}$(teacher, \nabla L, Adam)$
  \EndWhile
  \State \Return $teacher$
\EndFunction

\vspace{0.4em}

\Function{FeatureUnlearn}{model, forgetFeat}
  \State $(W_u, b_u) \gets model.\textproc{GetWeights}(\texttt{first\_dense},$
  \Statex \hspace{2.6cm}$\textproc{Index}(forgetFeat))$
  \State $W_v \gets 0.0$; $b_v \gets 0.0$
  \State $model \gets model.\textproc{UpdateWeights}(\texttt{first\_dense},$
  \Statex \hspace{2.6cm}$\textproc{Index}(forgetFeat), W_v, b_v)$
  \State \Return model
\EndFunction

\vspace{0.4em}

\Function{DistillToStudent}{teacher, df\_train, $T$} 
  \State $student \gets \textproc{A2C()}$
  \State $cls\_weights \gets \textproc{ComClsWeights}(df\_train)$
  \State $t\_probs \gets \textproc{Softmax}(teacher(df\_train)/T)$
  \While{NOT \textproc{Converged}(student, EarlyStop)}
    \State $(x, y) \gets \textproc{Sample}(df\_train)$
    \State $L \gets \alpha \cdot \textproc{CE}(student(x), y)$
    \Statex \hspace{1.5cm}$+ (1-\alpha) \cdot T^2 \cdot \textproc{KL}(student(x)/T,\ t\_probs)$
    \Statex \hspace{1.5cm}$+ \textproc{MSE}(critic(x), y)$
    \Statex \hspace{1.5cm}$+ \lambda_{attn} \cdot \textproc{Entropy}(attention)$
    \State \textproc{Step}(student, $\nabla L$, Adam)
  \EndWhile
  \State \Return $student$
\EndFunction

\vspace{0.4em}

\Function{RestoreFeatures}{model, forgetFeat}
  \State $(W_v, b_v) \gets \textproc{SavedWeightsBiases}(\texttt{first\_dense},$
  \Statex \hspace{2.6cm}$\textproc{Index}(forgetFeat))$
  \State $model \gets model.\textproc{UpdateWeights}(\texttt{first\_dense},$
  \Statex \hspace{2.6cm}$\textproc{Index}(forgetFeat), W_v, b_v)$
  \State \Return model
\EndFunction

\vspace{0.4em}

\State $df_{tr}, df_{val}, df_{te} \gets \Call{Preprocess}{df}$
\State $teacher \gets \Call{TrainA2CTeacher}{df_{tr}}$
\State $student \gets \Call{DistillToStudent}{teacher,\ df_{tr},\ T}$
\vspace{0.2em}
\State $base\_models (\mathcal{B}) \gets [teacher,\ student]$
\State $unlearn\_models (\mathcal{UL}) \gets [\,]$
\State $restore\_models (\mathcal{RE}) \gets [\,]$
\vspace{0.2em}
\ForAll{$m \in base\_models$}
  \State $unlearn\_models \gets \Call{FeatureUnlearn}{m,\ df_{fg}}$
\EndFor
\vspace{0.2em}
\ForAll{$m\_ul \in unlearn\_models$}
  \State $restore\_models \gets \Call{RestoreFeatures}{m\_ul,\ df_{fg}}$
\EndFor
\vspace{0.2em}
\State $all\_models \gets\mathcal{B} \cup \mathcal{UL} \cup \mathcal{RE}$

\ForAll{$x \in df_{te}$}
  \ForAll{$m \in all\_models$}
    \State $\hat{y} \gets \textproc{Predict}(m,\ x)$
    \State $\Call{ExplainWithLIME}{\hat{y}}$
  \EndFor
\EndFor
\State $\textproc{CompareResults}(all\_models)$

\end{algorithmic}
\end{minipage}}
\label{pseudocode}
\end{algorithm}

To conduct the research, a high-performance workstation was configured with 64GB of RAM, a 3.40GHz processor, and an NVIDIA 12GB GPU running Windows 11. This setup ensured optimal processing and memory capacity for data-intensive tasks.

\subsection{Model Evaluation}
In this research, we tested the model's performance using various commonly used metrics, such as accuracy, F1 score, and loss.

Regardless of class, accuracy determines the overall accuracy of the model's predictions and evaluates its performance.

\begin{equation}\label{eq:6}
\textit{Accuracy} = \frac{\textit{TP} + \textit{TN}}{\textit{TP} + \textit{TN} + \textit{FP} + \textit{FN}}
\end{equation}

Precision measures the percentage of the model's predictions that are correct, and recall measures the percentage of relevant data points that the model correctly identified. The mean of precision and recall is the F1 score.

\begin{equation}\label{eq:7}
\textit{Precision} = \frac{\textit{TP}}{\textit{TP} + \textit{FP}}
\end{equation}

\begin{equation}\label{eq:8}
\textit{Recall} = \frac{\textit{TP}}{\textit{TP} + \textit{FN}}
\end{equation}

\begin{equation}\label{eq:9}
\textit{F1 score} = \frac{\textit{2*Precision*Recall}}{\textit{Precision} + \textit{Recall}}
\end{equation}

In equations \ref{eq:6}-\ref{eq:9}, $TP$ denotes true positives, $TN$ denotes
true negatives, $FP$ denotes false positives, and $FN$ denotes
false negatives.

Another evaluation metric, Binary Cross-Entropy ($BCE$) is a loss function used in this paper for binary classification. It measures the difference between accurate labels ($y$) and predicted probabilities ($\hat{y}$). It penalizes inaccurate predictions more when predictions are confident but wrong.

\begin{equation}\label{eq:10}
\text{BCE} = - \frac{1}{n} \sum_{i=1}^{n} \left[ y_i \log(\hat{y}_i) + (1 - y_i) \log(1 - \hat{y}_i) \right]
\end{equation}

In the equation \ref{eq:10}, $y_i$ is the actual label, $\hat{y}_i$ is the predicted probability, and $n$ is the number of samples.

\section{Result and Analysis}\label{result_and_analysis}
This section includes a detailed explanation of the results (only 25\% datapoints) obtained using the algorithms and dataset described in the methodology section. To ensure transparency, we used XAI techniques like LIME to explain how each model makes predictions and included two sample test case scenarios. Finally, we will evaluate the study's shortcomings and explore potential areas for improvement.

Table-\ref{table_result} summarizes the Knowledge Distillation framework's performance across three stages: before feature removal, after feature removal, and after feature restoration. We conducted an evaluation based on accuracy, F1 Score, and loss for A2C, Q-Learning, and Meta-RL algorithms on 25\% and 30\% data subsets, comparing both Teacher and Student models. Before feature removal, Meta-RL delivers the best Teacher performance, achieving 99.704\% accuracy and 99.852\% F1 on the 30\% subset. For Student models, A2C leads with 99.602\% accuracy and 99.800\% F1 on the 25\% subset. Q-Learning performs well but slightly trails the other two. After feature removal, performance drops for some algorithms, particularly Meta-RL on the 30\% subset, where Teacher's accuracy drops to 85.336\% and Student accuracy to 79.471\%, showing its sensitivity to missing features. A2C remains relatively stable, with Student accuracy at 99.354\% (25\% subset) and 93.163\% (30\% subset). Q-Learning experiences moderate drops, with Student accuracy falling to 98.109\% on the 25\% subset. After feature restoration, most models regain near-original performance. Meta-RL Teacher scores return to 99.704\% accuracy and 99.852\% F1, while A2C Student scores hit 99.603\% accuracy and 99.801\% F1 on the 25\% subset. Overall, A2C proves the most stable across all stages, while Meta-RL performed better for the Teacher model only, particularly with limited datapoints.

In Table-\ref{table_result}, we present the results obtained from multiple algorithms (A2C, Q-learning, and Meta-RL) explored in this research. However, our detailed discussion focuses on the outcomes and behavior of A2C.


\begin{table*}[]
\caption{Performance comparison of the Knowledge Distillation framework across three stages-(i) baseline with all features, (ii) after feature removal to simulate unlearning, and (iii) after feature restoration, utilizing A2C, Q-Learning, and Meta-RL}
\label{table_result}
\renewcommand{\arraystretch}{1.5}
\centering
\resizebox{\textwidth}{!}{%
\begin{tabular}{ccccccccccccc}
\hline \hline
 & & & \multicolumn{3}{c}{\textbf{Before Feature Removal}} & \multicolumn{3}{c}{\textbf{After Feature Removal}} & \multicolumn{3}{c}{\textbf{After Feature Restoration}} \\ \hline
\textbf{Algorithm} & \textbf{Data(\%)} & \textbf{Model} & \textbf{Accuracy(\%)} & \textbf{F1 Score(\%)} & \textbf{Loss} & \textbf{Accuracy(\%)} & \textbf{F1 Score(\%)} & \textbf{Loss} & \textbf{Accuracy(\%)} & \textbf{F1 Score(\%)} & \textbf{Loss} \\ \cline{1-12}
\multirow{4}{*}{\textbf{A2C}} & \multirow{2}{*}{25} & Teacher & 99.599 & 99.799 & 0.056 & 99.310 & 99.654 & 0.018 & 99.558 & 99.778 & 0.011 \\ \cline{3-12}
                   &                    & Student & \textbf{99.602} & 99.800 & 0.025 & 99.354 & 99.676 & 0.013 & 99.603 & 99.801 & 0.009 \\ \cline{2-12}
                   & \multirow{2}{*}{30} & Teacher & 99.471 & 99.735 & 0.146 & 99.388 & 99.693 & 0.052 & 99.471 & 99.735 & 0.038 \\ \cline{3-12}
                   &                    & Student & 93.314 & 96.540 & 0.048 & 93.163 & 96.459 & 0.132 & 93.314 & 96.540 & 0.095 \\ \cline{1-12}
\multirow{4}{*}{Q-Learning} & \multirow{2}{*}{25} & Teacher & 99.593 & 99.796 & 0.033 & 99.240 & 99.618 & 0.071 & 99.480 & 99.739 & 0.057 \\ \cline{3-12}
                   &                    & Student & 99.240 & 99.618 & 0.071 & 98.109 & 99.045 & 0.136 & 89.797 & 99.395 & 0.114 \\ \cline{2-12}
                   & \multirow{2}{*}{30} & Teacher & 99.667 & 99.833 & 0.034 & 99.456 & 99.727 & 0.046 & 99.656 & 99.828 & 0.034 \\ \cline{3-12}
                   &                    & Student & 99.456 & 99.727 & 0.046 & 99.181 & 99.588 & 0.052 & 99.295 & 99.646 & 0.040 \\ \cline{1-12}
\multirow{4}{*}{Meta-RL} & \multirow{2}{*}{25} & Teacher & 99.687 & 99.843 & 0.020 & 99.445 & 99.722 & 0.012 & 99.687 & 99.843 & 0.008 \\ \cline{3-12}
                   &                    & Student & 99.019 & 99.507 & 0.048 & 98.912 & 99.453 & 0.069 & 99.019 & 99.507 & 0.044 \\ \cline{2-12}
                   & \multirow{2}{*}{30} & Teacher & 99.704 & 99.852 & 0.024 & 85.336 & 92.084 & 0.333 & 99.704 & 99.852 & 0.007 \\ \cline{3-12}
                   &                    & Student & 98.931 & 99.463 & 0.067 & 79.471 & 88.556 & 0.125 & 98.931 & 99.463 & 0.039 \\ \hline \hline
\end{tabular}%
}
\end{table*}
\subsection{Reinforcement Learning Results} 
\begin{figure}[htbp]
  \centering
  \includegraphics[width=0.9\columnwidth]{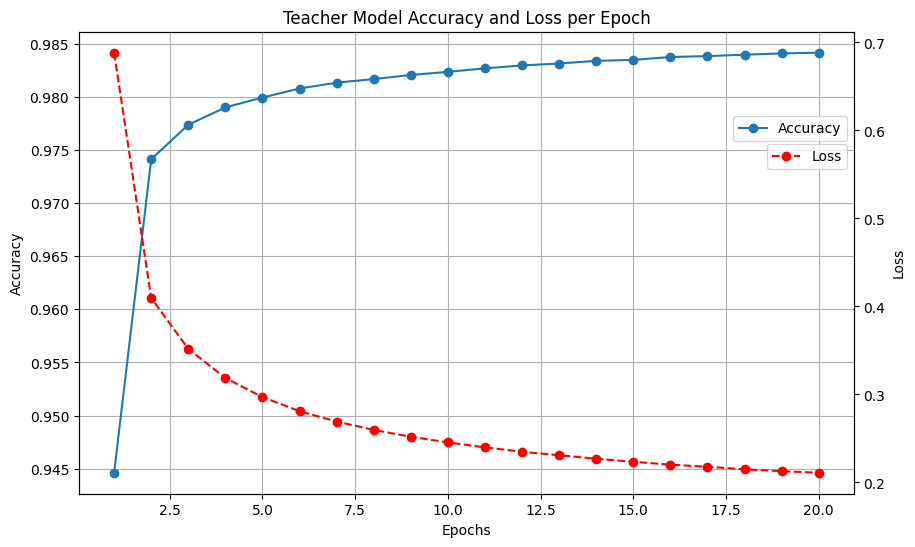}
  \caption{Training accuracy and loss at each epoch of Teacher model}
  \label{fig5}
\end{figure}

\begin{figure}[htbp]
  \centering
  \includegraphics[width=0.8\columnwidth]{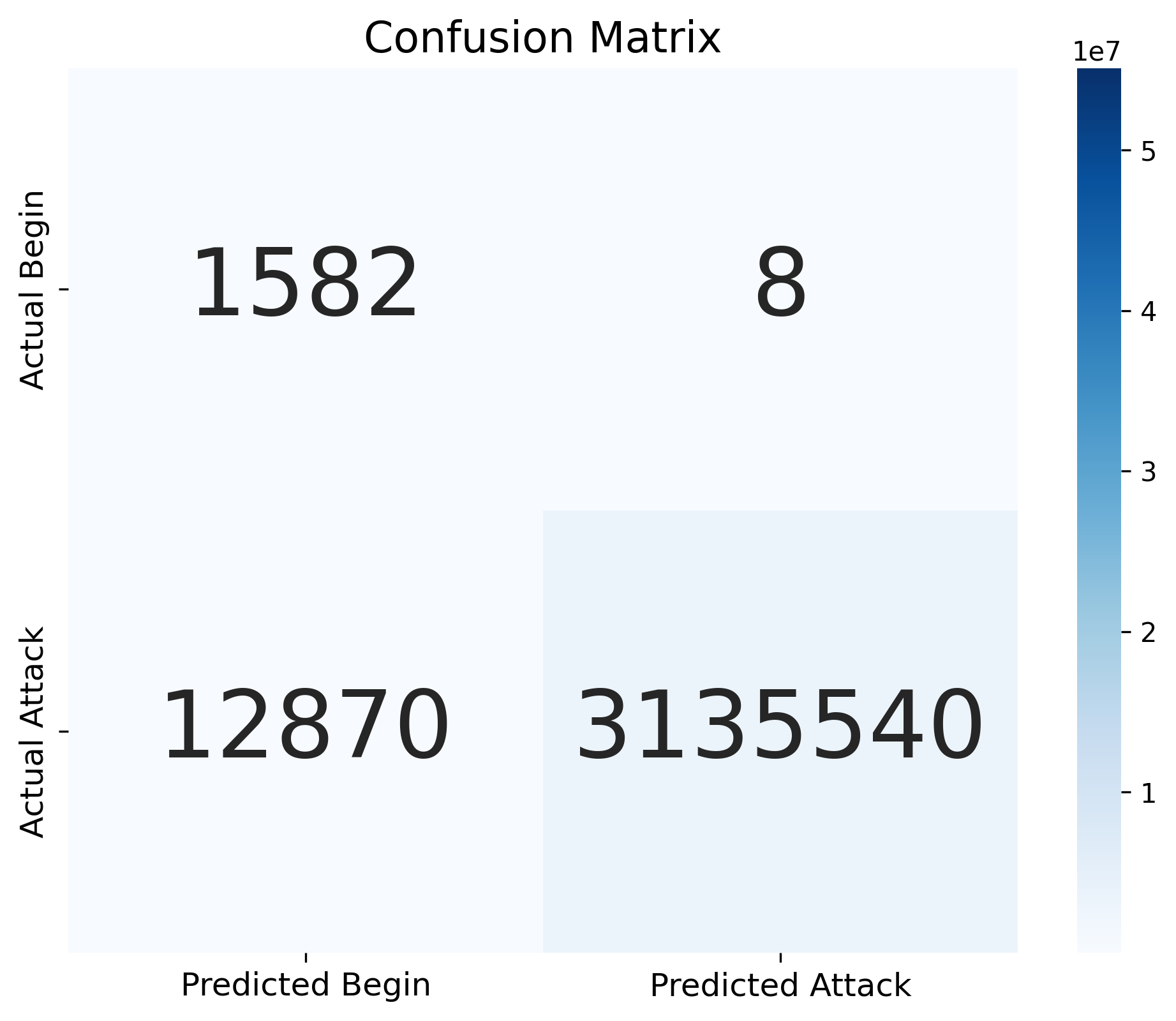}
  \caption{Confusion matrix of Teacher model}
  \label{fig5.1}
\end{figure}

The teacher model demonstrated consistent performance improvement during its 20-epoch training process in the knowledge distillation framework. Initially, it achieved a training accuracy of 88.72\% with a total loss of 1.2470 and a remarkably high validation accuracy of 99.60\%, indicating strong generalization from the start. The model stabilized around 98.41\% training accuracy as training progressed, with a final loss of 0.2115 by epoch 20. Notably, in epoch 4, the validation loss dropped to 0.1869, and by epoch 10, it reached a low of 0.1410, showing stable learning with minimal overfitting. The final evaluation of the test set revealed an impressive accuracy of 99.59\%, an F1 score of 99.79\%, and a minimal loss of 0.0534, confirming the model’s robustness and high precision-recall balance. In Fig-\ref{fig5}, we visualized the training accuracy and loss over epochs to illustrate the model’s convergence. The confusion matrix in Fig-\ref{fig5.1} provides deeper insight into the model’s classification performance across classes.

\begin{figure}[htbp]
  \centering
  \includegraphics[width=0.9\columnwidth]{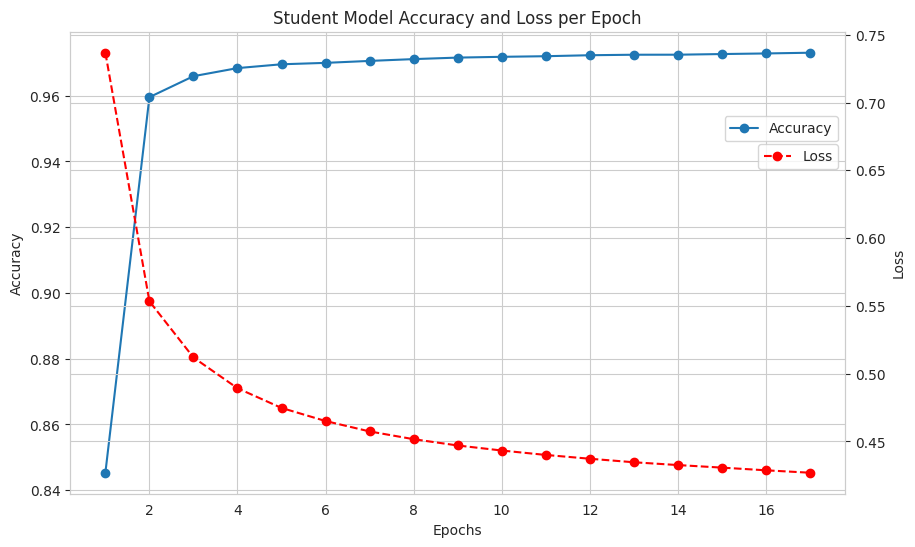}
  \caption{Training accuracy and loss at each epoch of Student model}
  \label{fig6}
\end{figure}

\begin{figure}[htbp]
  \centering
  \includegraphics[width=0.8\columnwidth]{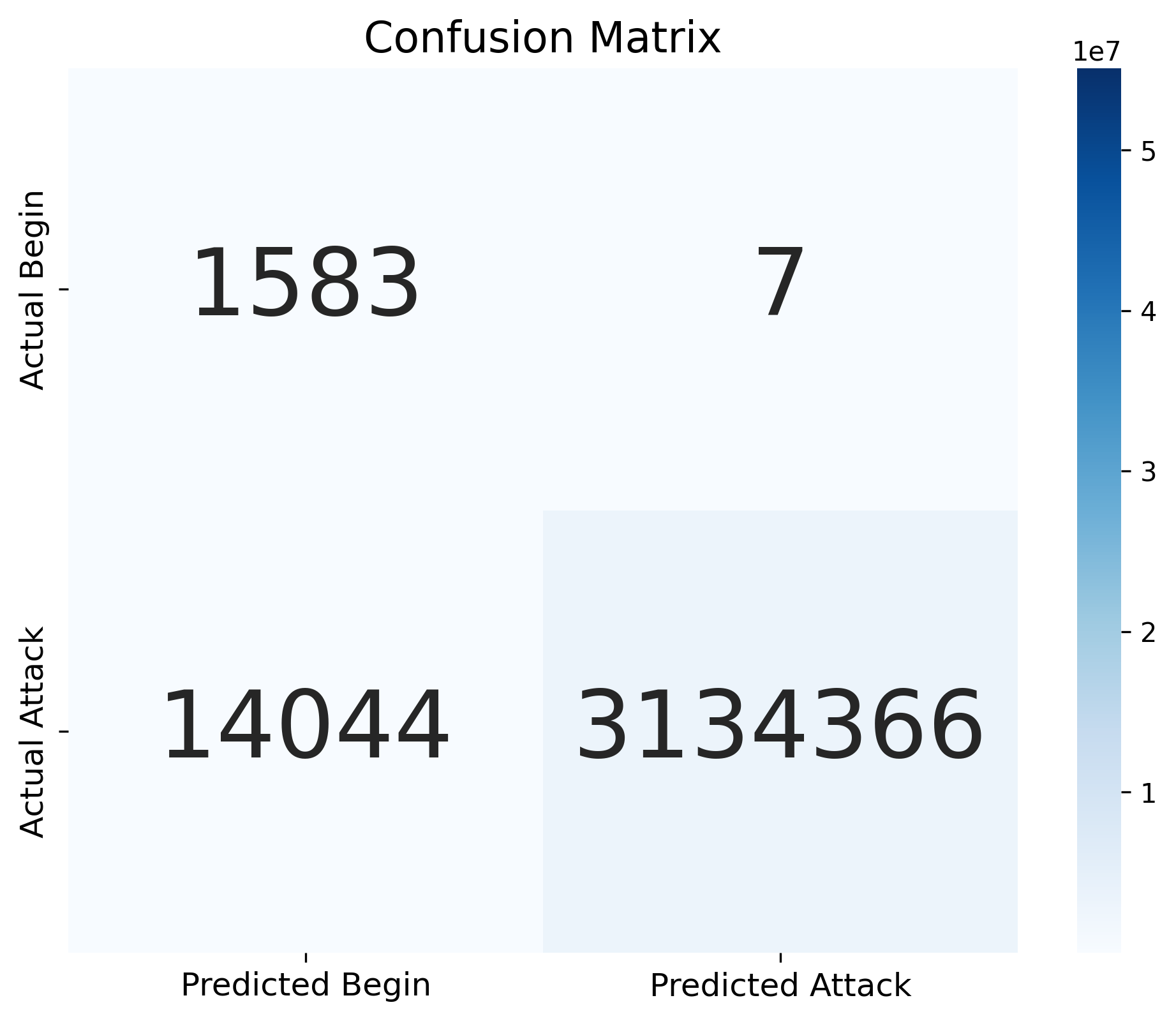}
  \caption{Confusion matrix of Student model}
  \label{fig6.1}
\end{figure}

The student model was trained over 17 epochs and exhibited consistent performance improvements throughout the training process. In the first epoch, it attained a training accuracy of 74.50\% with a total loss of 1.0584, while the validation accuracy was already remarkably high at 99.61\%, reflecting strong initial generalization. As training progressed, the model’s accuracy steadily improved, converging to a stable level of 97.31\% with a final loss of 0.4271 by epoch 17. The validation metrics also demonstrated reliable convergence, with the validation loss reaching its lowest value of 0.3350 in the final epoch, confirming minimal overfitting. Figure-\ref{fig6} illustrates the convergence trends of training and validation accuracy and loss across epochs. In the final evaluation, the student model achieved an impressive accuracy of 99.55\%, an F1 score of 99.78\%, and a minimal test loss of 0.0444. These results highlight the robustness and effectiveness of the knowledge distillation framework, where the student model successfully approximates the teacher model’s performance while maintaining a strong precision–recall balance. For further performance insights, the confusion matrix of the student model is provided in Fig-\ref{fig6.1}, offering a detailed view of class-level classification outcomes.

\subsection{Feature Unlearning Results}
In this study, an unlearning technique was used to maintain data security. We first applied the unlearning process to the teacher and student models, which we named teacher unlearning and student unlearning, respectively.

\begin{figure}[htbp]
  \centering
  \includegraphics[width=0.9\columnwidth]{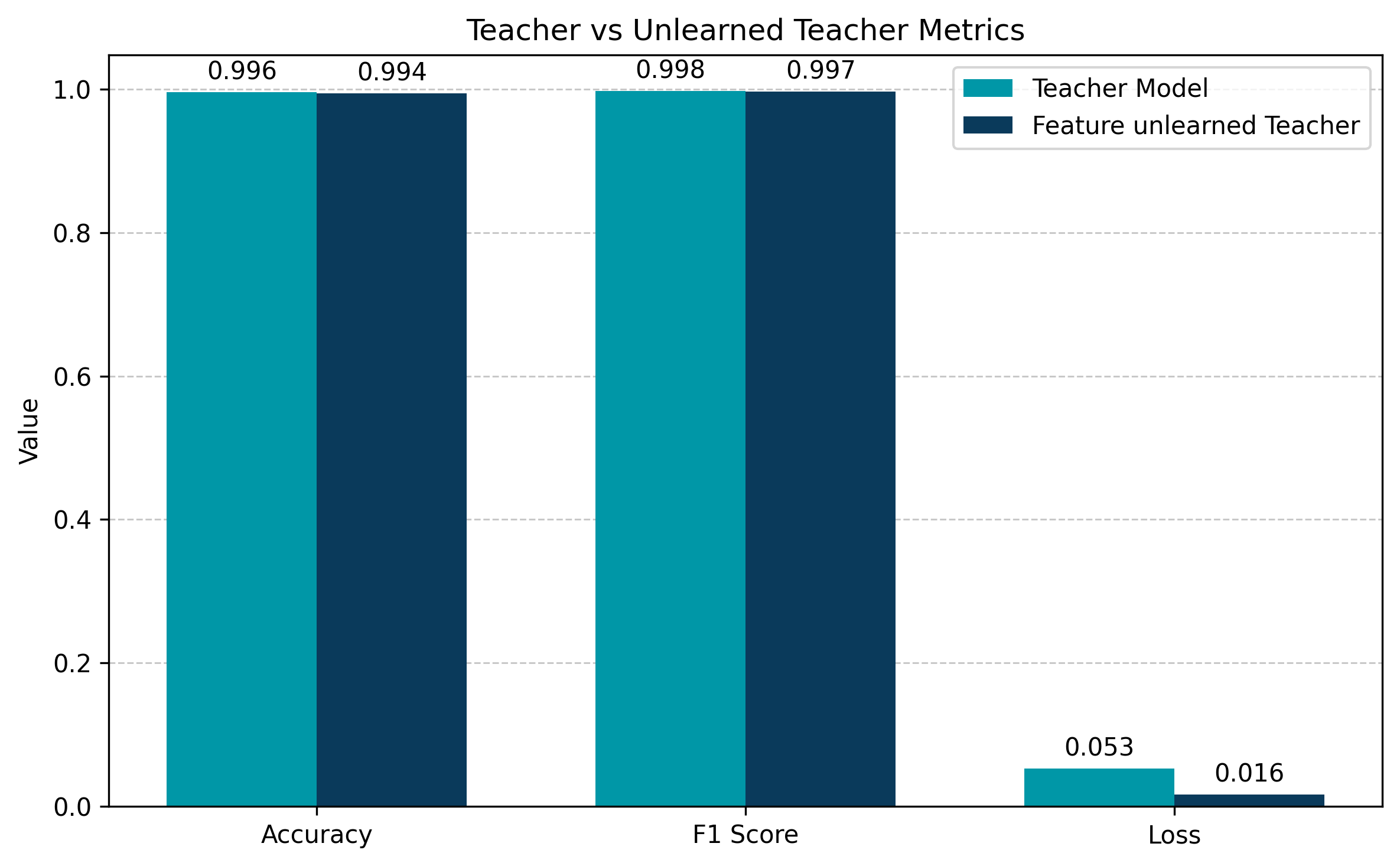}
  \caption{Comparison of Teacher and Teacher unlearn models: accuracy, loss and f1 score}
  \label{fig7}
\end{figure}

\begin{figure}[htbp]
  \centering
  \includegraphics[width=0.8\columnwidth]{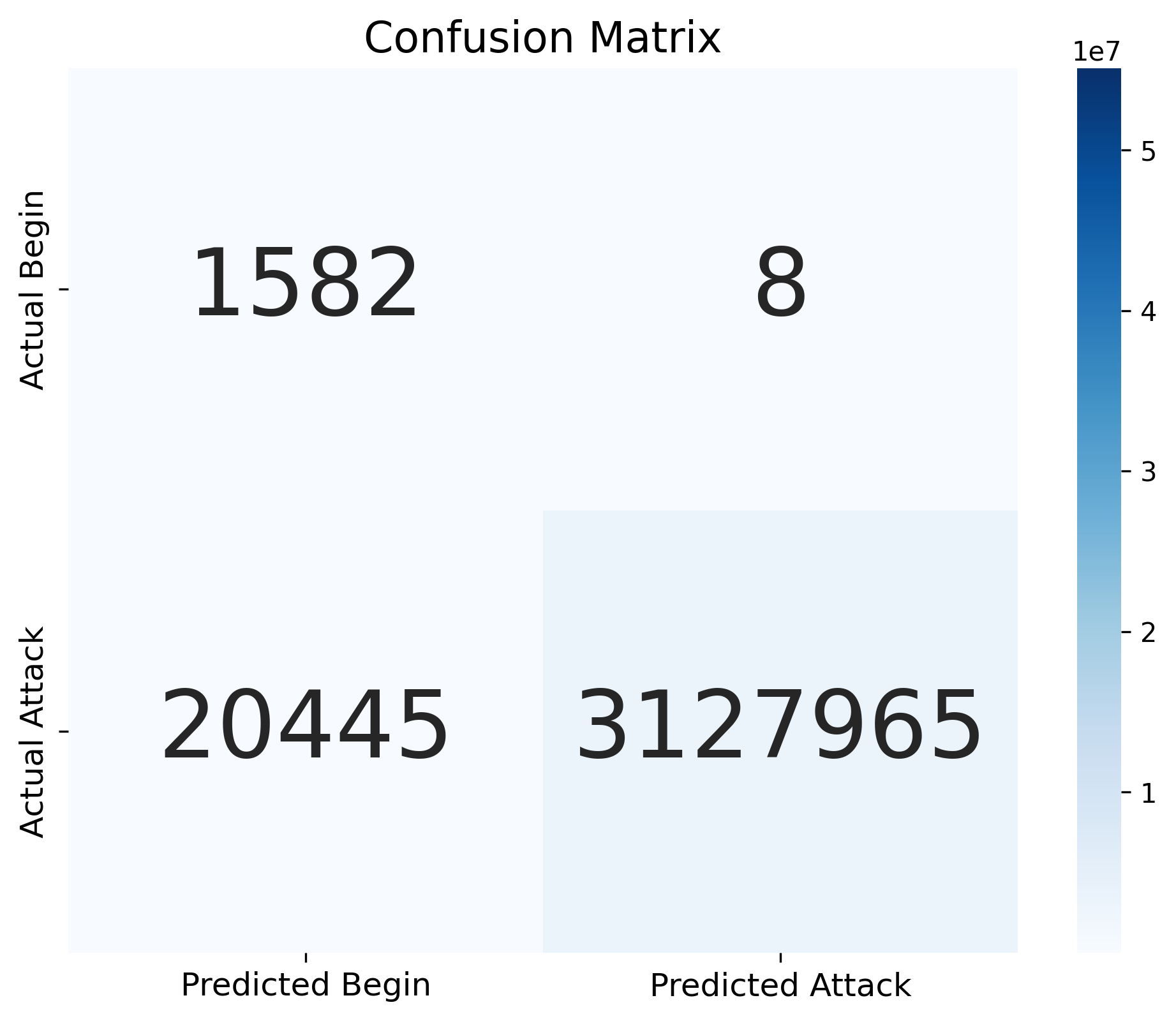}
  \caption{Confusion matrix of Teacher unlearn model}
  \label{fig7.1}
\end{figure}

To strengthen data security, we applied post-hoc weight modification to the first layer of the teacher model. This adjustment ensures the model no longer relies on the removed feature, \textit{flgs} when making predictions. To evaluate the impact of this unlearning process, we re-assessed the model’s performance using accuracy and F1-score. The results showed that the unlearned teacher model achieved an accuracy of 99.35\% and an F1 score of 99.67\%, with a minimal loss of 0.0156. Unlike previous unlearning attempts that involved highly influential features, the limited performance drop, in this case, suggests that \textit{flgs} had a relatively smaller impact on overall predictions. Nonetheless, this outcome confirms that the model effectively eliminated reliance on the targeted feature. This method improves both model interpretability and data security while maintaining high performance. We have visualized the comparison between the original and unlearned teacher models in Fig-\ref{fig7}. Additionally, to provide visibility of class-wise prediction behaviour, we generated the confusion matrix of the unlearned teacher model, as shown in Fig-\ref{fig7.1}.

\begin{figure}[htbp]
  \centering
  \includegraphics[width=0.9\columnwidth]{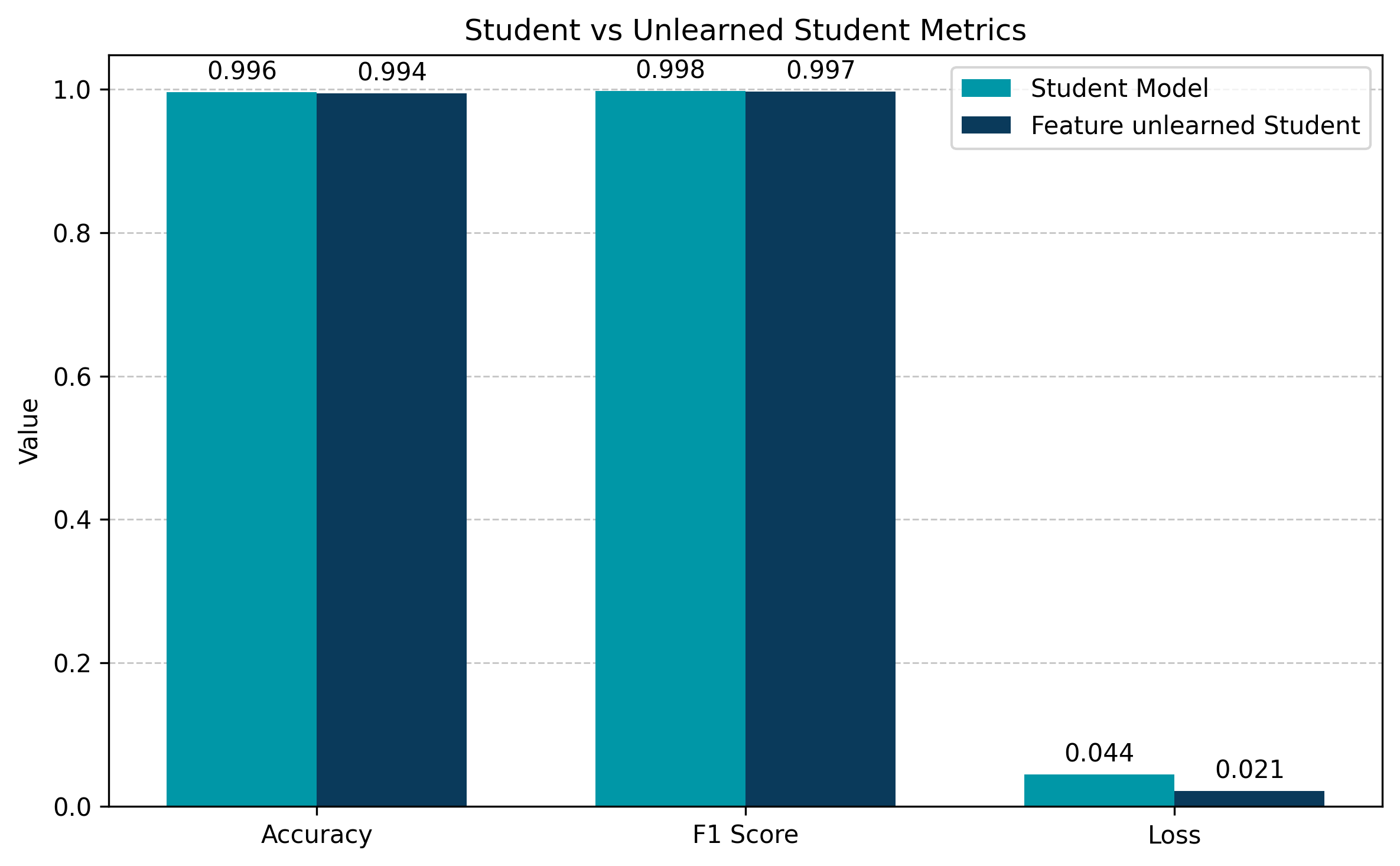}
  \caption{Comparison of Student and Student unlearn models: accuracy, loss and f1 score}
  \label{fig8}
\end{figure}

\begin{figure}[htbp]
  \centering
  \includegraphics[width=0.8\columnwidth]{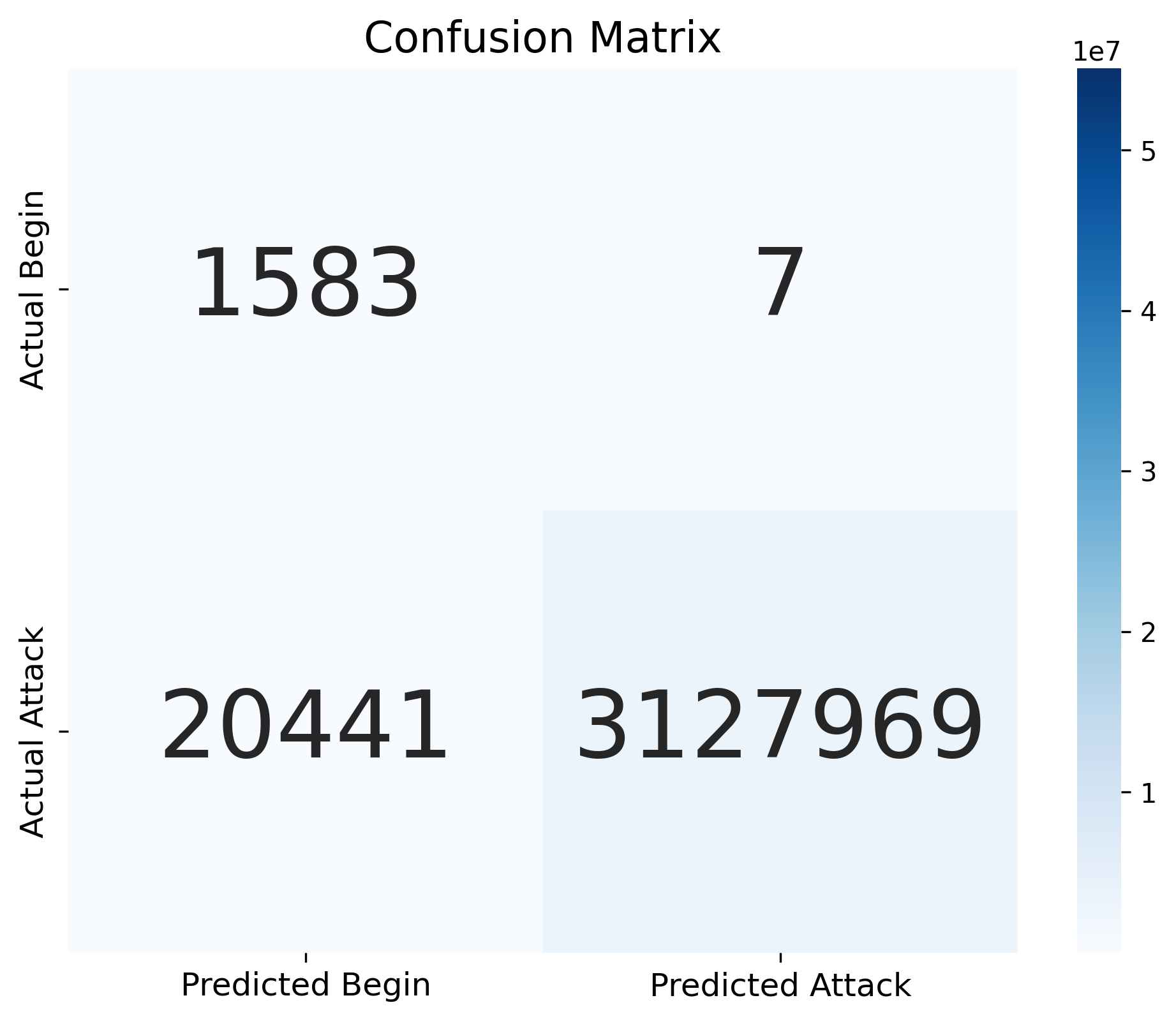}
  \caption{Confusion matrix of Student unlearn model}
  \label{fig8.1}
\end{figure}

To improve data security in the student model, we applied post-hoc weight modification to the first layer of the network. To evaluate the impact of this adjustment, we re-assessed the model using accuracy and F1-score. The results showed that the unlearned student model achieved a test accuracy of 99.35\% and an F1 score of 99.67\%, with a total loss of 0.0213. The minimal decrease in both metrics indicates that the model successfully unlearned the influence of the removed feature \textit{flgs}, which had a relatively limited effect on its decision-making. This unlearning approach maintains high performance while improving interpretability and data privacy. We have visualized the performance comparison between the original and unlearned student models in figure-\ref{fig8}. Additionally, to highlight the class-wise behaviour of the unlearned model, we generated its confusion matrix as shown in fig-\ref{fig8.1}.

\begin{figure}[htbp]
  \centering
  \includegraphics[width=0.9\columnwidth]{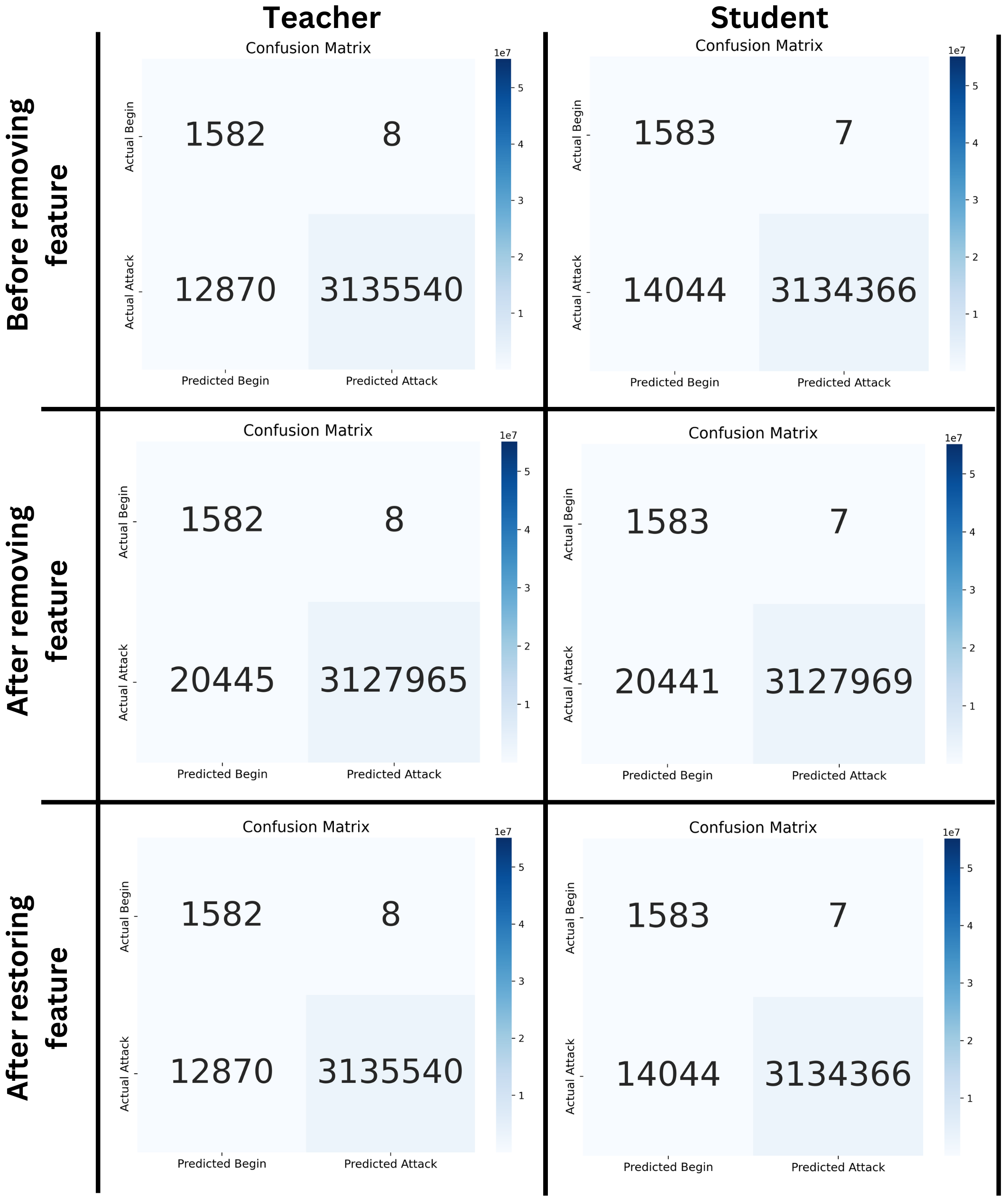}
  \caption{Confusion matrices for all three stages: before removing the feature, after removing the feature, and after restoring the feature}
  \label{total_con}
\end{figure}

Nevertheless, to check the robustness of the overall framework, we restored the feature flags we had previously removed and reevaluated both the teacher and student models. After restoring the original weights and biases, the teacher model reached an accuracy of 99.59\% and an F1 score of 99.79\%, with a very low loss of 0.0099. The student model also performed similarly well, with an accuracy of 99.55\%, an F1 score of 99.77\%, and a loss of 0.0128. These results show that the unlearning process is fully reversible and does not cause any permanent drop in performance. To clarify this, we visualized the confusion matrices for all three stages: before removing the feature, after removing the feature, and after restoring the feature; all together in one frame, as shown in fig-\ref{total_con}.

\subsection{XAI: Teacher, Student and Post‑Unlearning}
We have applied Explainable AI (XAI) using LIME (Local Interpretable Model-agnostic Explanations) for the model's transparency and visualized which features impacted the most in making such a prediction. We have selected test instance 100 and this instance is an attack traffic.

\begin{figure}[htbp]
  \centering
  \includegraphics[width=0.8\columnwidth]{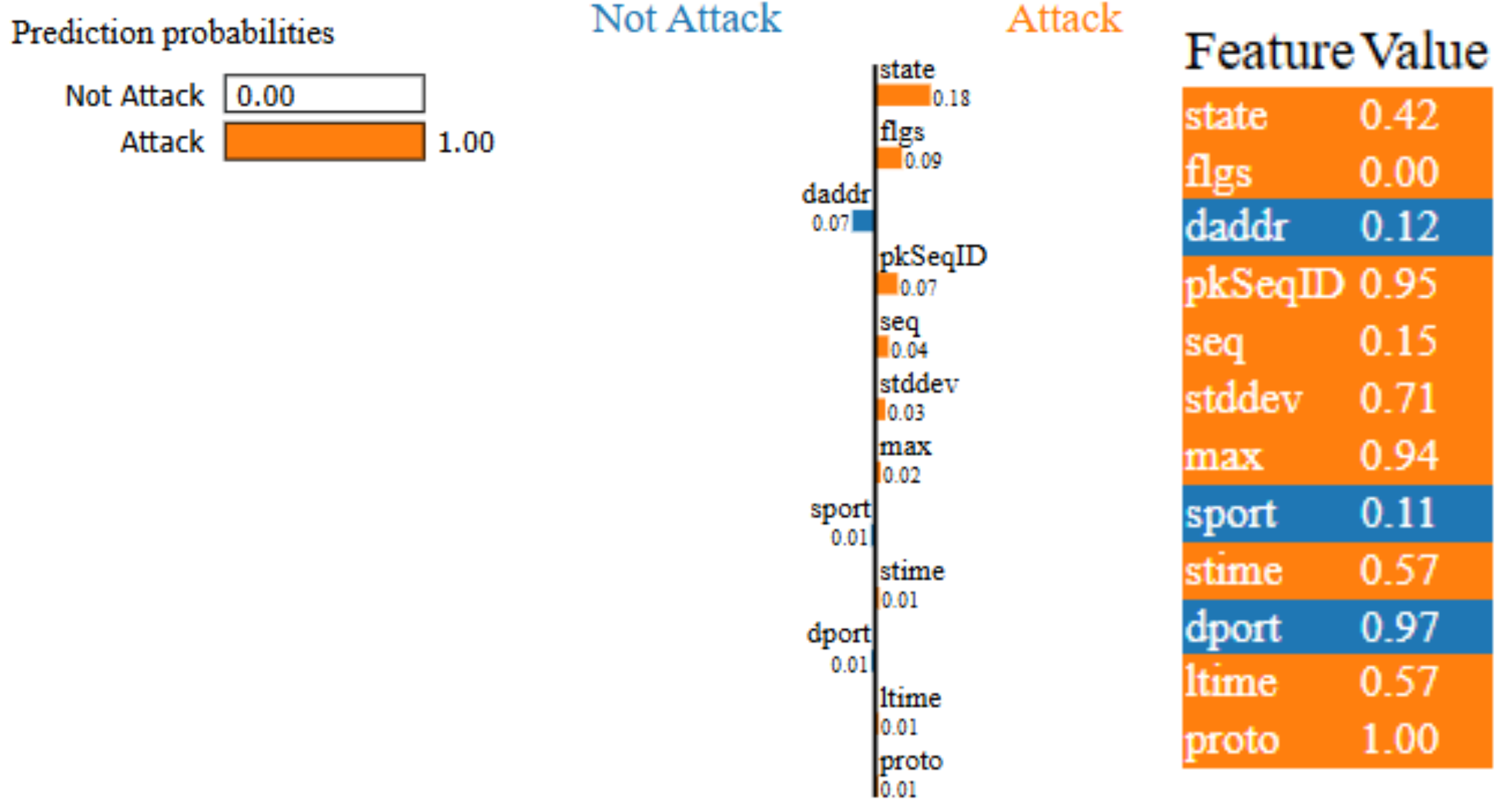}
  \caption{XAI interpretation of the Teacher model’s prediction for an attack case}
  \label{fig9}
\end{figure}

\begin{figure}[htbp]
  \centering
  \includegraphics[width=0.9\columnwidth]{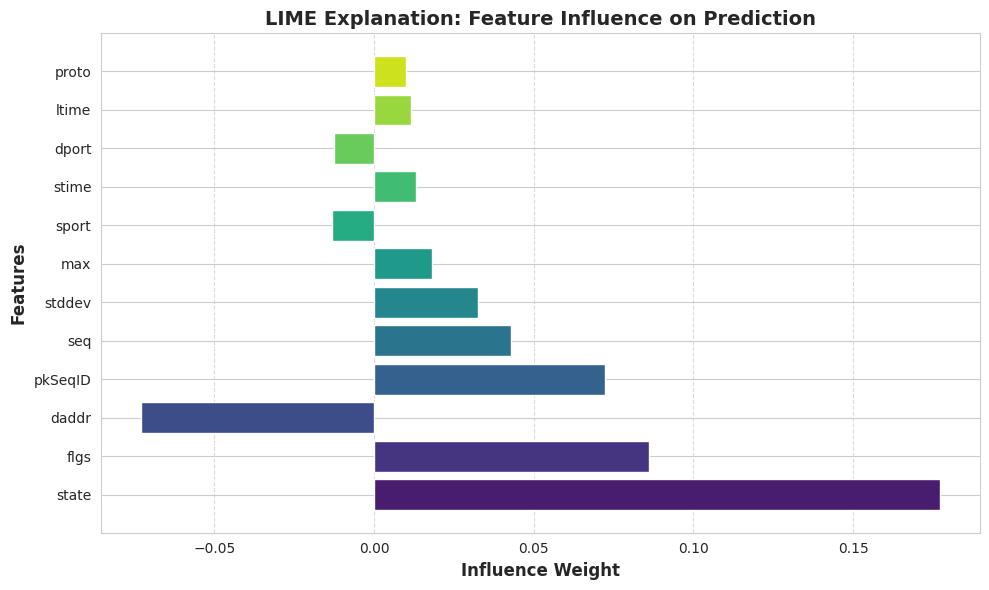}
  \caption{XAI interpretation of top 12 features impacts for the Teacher model}
  \label{fig9.1}
\end{figure}

Using the Teacher model, we have visualized the test case in figure-\ref{fig9}. In the figure-\ref{fig9} illustrates that the teacher model identified test case 100 as an attack with a 100\% probability. The feature value highlights feature contributions, with the \textit{state} and \textit{flags} having the highest impact. Other features like daddr, seq, sport, ete., contribute a little, while most others have minimal effects. Additionally, in fig-\ref{fig9.1}, we have visualized the top 12 feature impacts.

\begin{figure}[htbp]
  \centering
  \includegraphics[width=0.8\columnwidth]{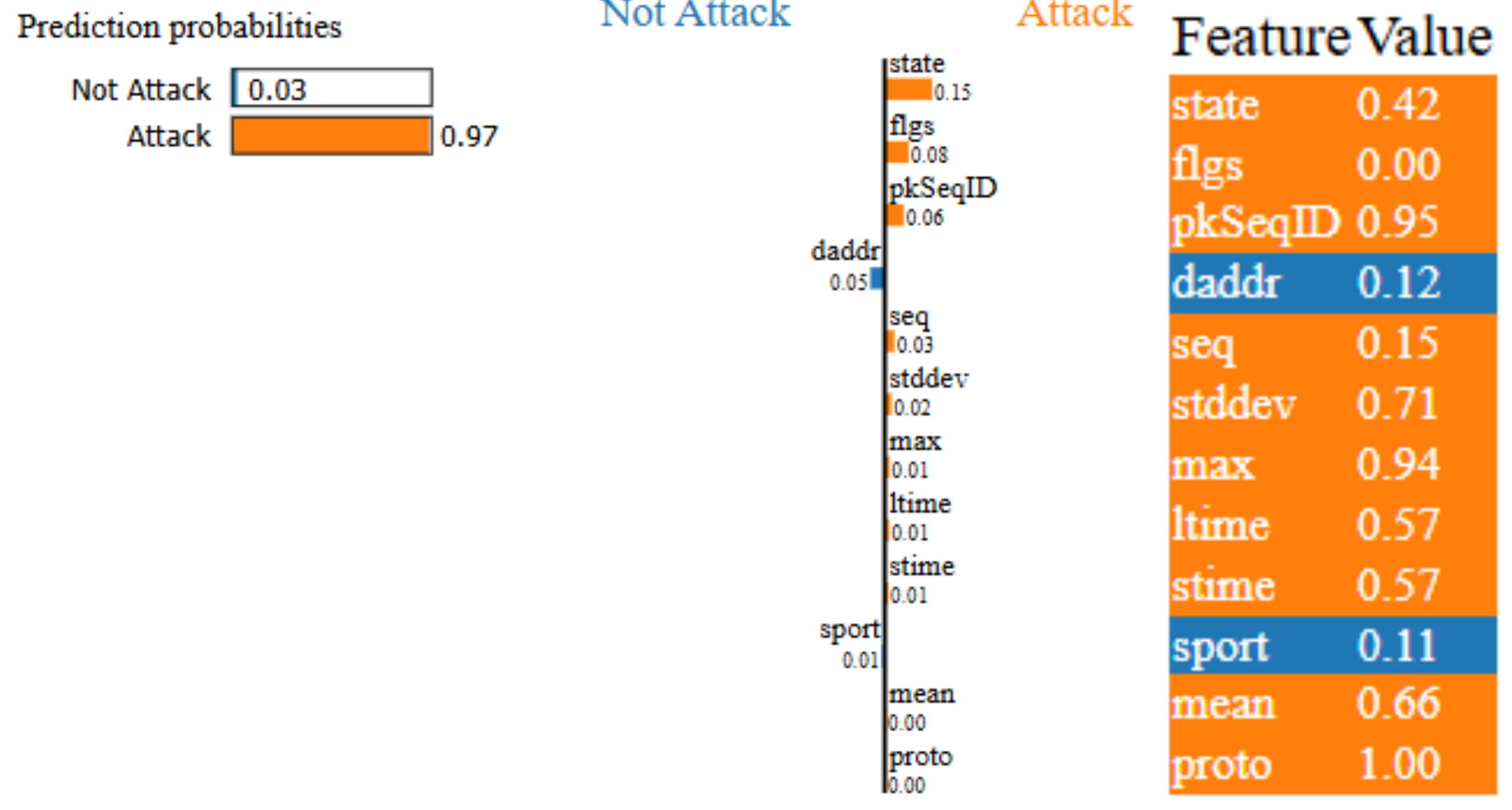}
  \caption{XAI interpretation of the Student model’s prediction for an attack case}
  \label{fig10}
\end{figure}

\begin{figure}[htbp]
  \centering
  \includegraphics[width=0.9\columnwidth]{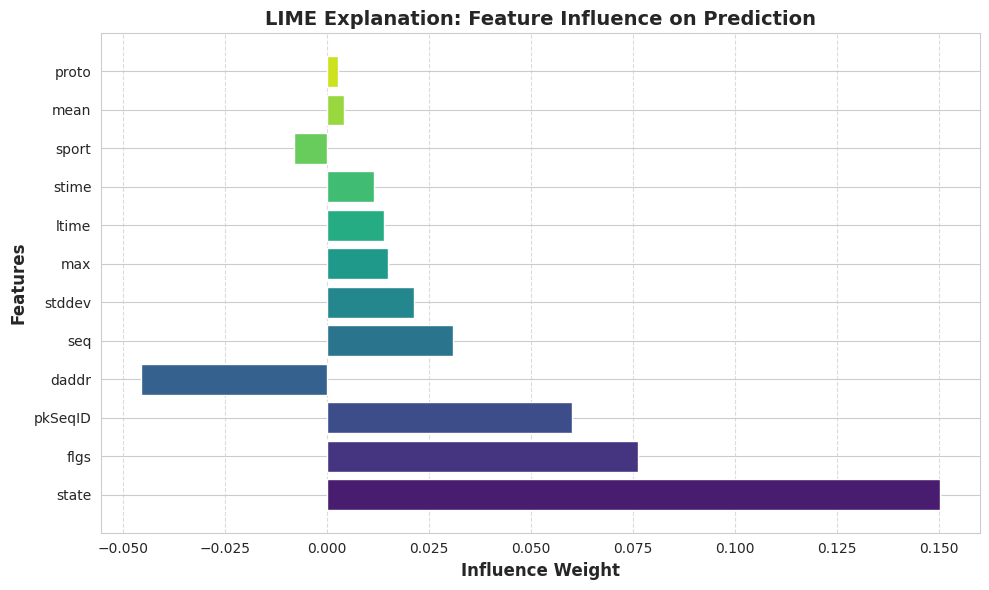}
  \caption{XAI interpretation of top 12 features impacts for the Student model}
  \label{fig10.1}
\end{figure}

We applied Lime on the Student model to show both test scenarios in figure-\ref{fig10}. Figure-\ref{fig10} shows that the model identified test case 100 as an attack with 97\% probability, with the \textit{state} and \textit{flags} having the highest impact. Other features, such as daddr, stddev, seq etc., make a minor contribution. In fig-\ref{fig10.1}, we have visualized the top 12 feature impacts of the Student model.

\begin{figure}[htbp]
  \centering
  \includegraphics[width=0.8\columnwidth]{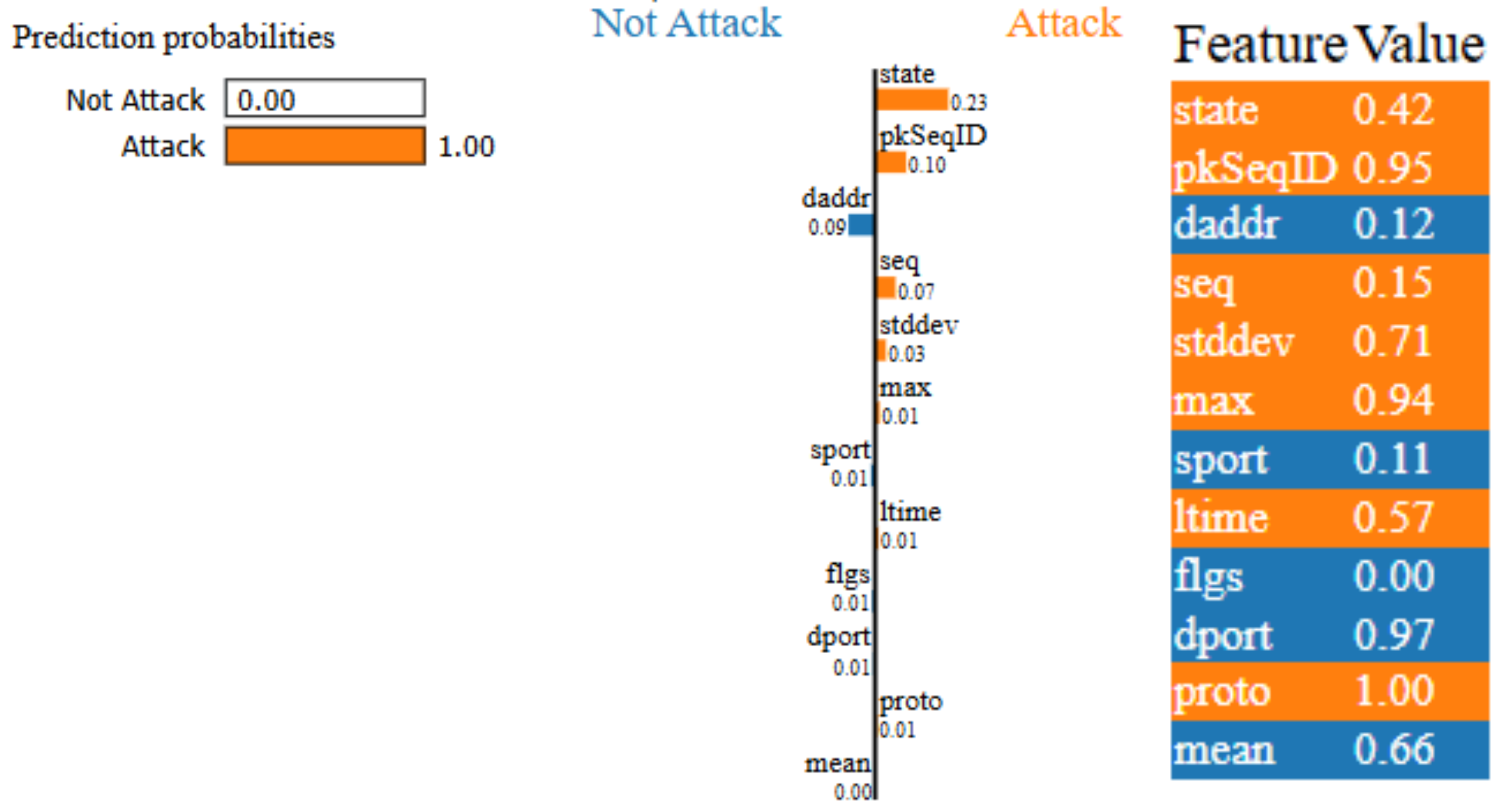}
  \caption{XAI interpretation of the Teacher Unlearn model’s prediction for an attack case}
  \label{fig11}
\end{figure}

\begin{figure}[htbp]
  \centering
  \includegraphics[width=0.9\columnwidth]{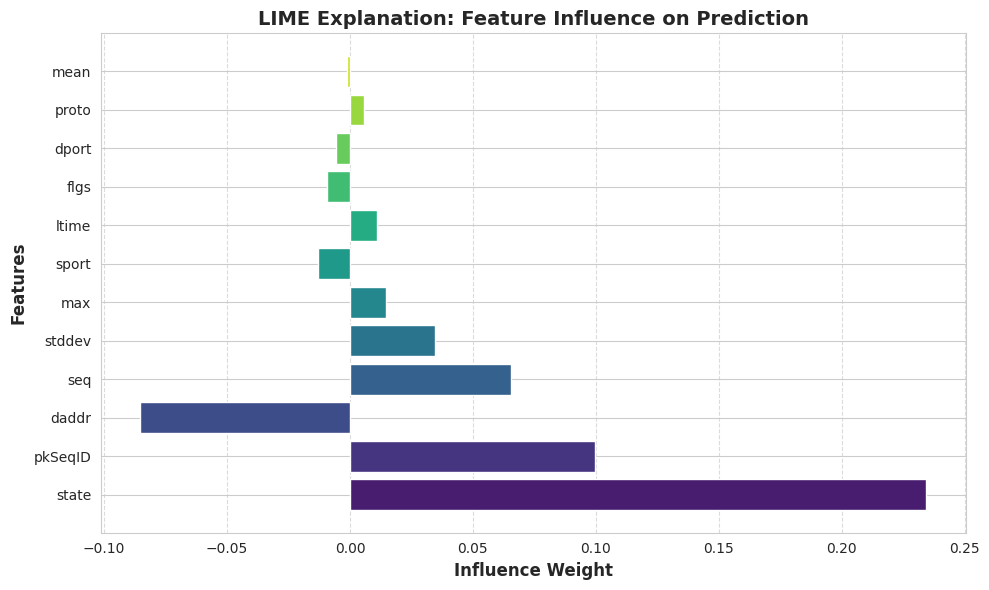}
  \caption{XAI interpretation of top 12 features impacts for the  Teacher Unlearn}
  \label{fig11.1}
\end{figure}

Figure-\ref{fig11} represents the prediction of test case-100 of the teacher unlearned model using LIME for interpretability. The model predicts a attack with a probability of 100\%. Feature contributions indicate that, \textit{state} and \textit{daddr} influenced the decision the most (as second most dominated feature remove by unlearning technique). However, their impact was not strong enough to alter the classification. Most other features contributed minimally, confirming that the unlearning process did not compromise the model's decision-making. In fig-\ref{fig11.1}, we have visualized the top 12 feature impacts of the Teacher Unlearn model.

\begin{figure}[htbp]
  \centering
  \includegraphics[width=0.8\columnwidth]{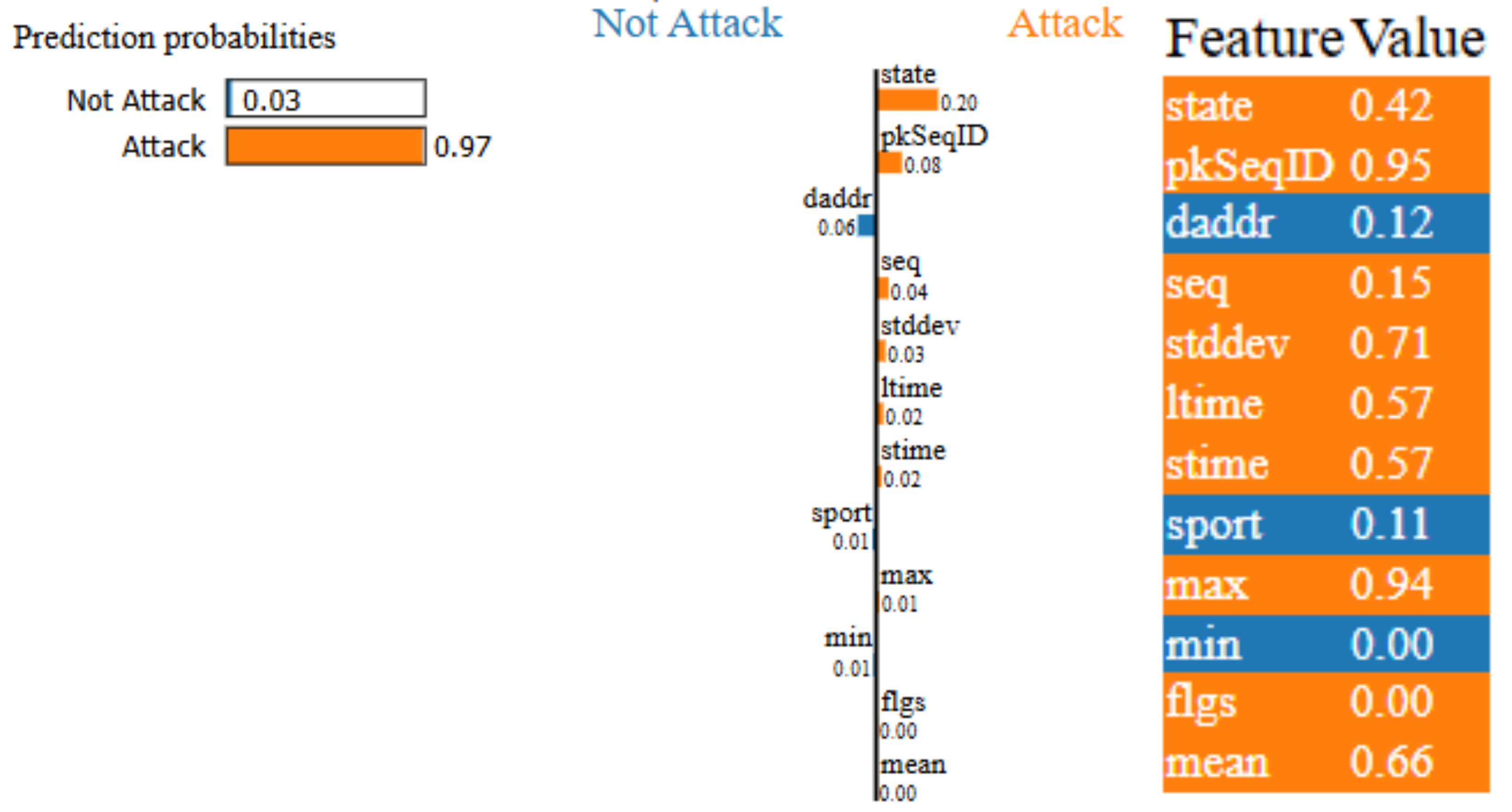}
  \caption{XAI interpretation of the Student Unlearn model’s prediction for an attack case}
  \label{fig12}
\end{figure}

\begin{figure}[htbp]
  \centering
  \includegraphics[width=0.9\columnwidth]{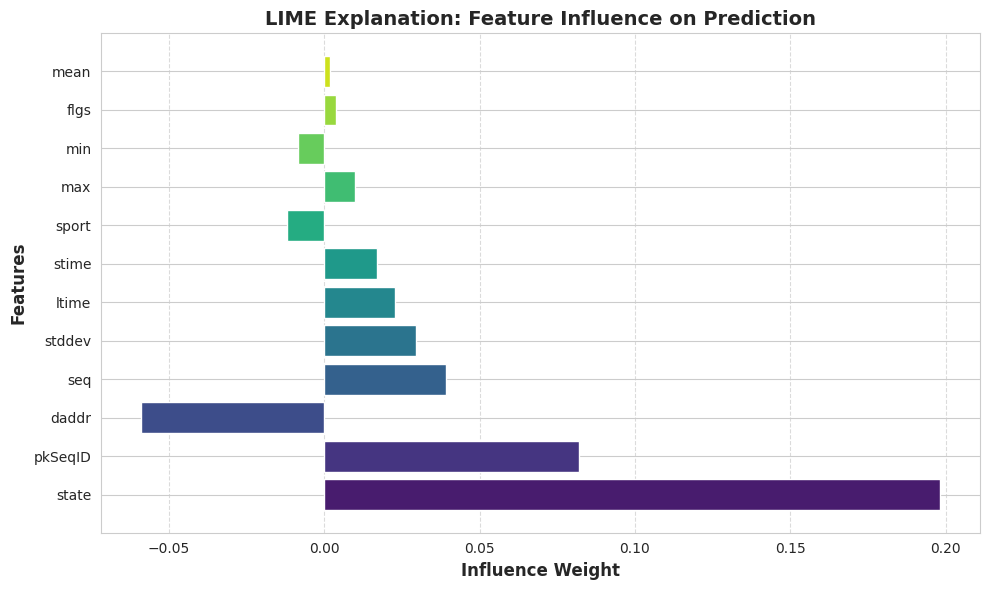}
  \caption{XAI interpretation of top 12 features impacts for the  Student Unlearn}
  \label{fig12.1}
\end{figure}

Figure-\ref{fig12} represents the prediction of test case-100 of the student unlearning model using LIME for interpretability. The model predicts a attack with a probability of 97\%. Feature contributions indicate that, \textit{state} and \textit{pkSeqID} influenced the decision the most (as second most dominated feature remove by unlearning technique). Other features contributed minimally, confirming that the unlearning process did not compromise the model's decision-making. In fig-\ref{fig12.1}, we have illiteratd  the top 12 feature impacts of the Student Unlearn model.

\subsection{Analysis}
This paper presented a compact reinforcement-learning-based botnet detection model, integrating Explainable AI (XAI) and feature unlearning for enhanced security. A key strength is its application of knowledge distillation, enabling the deployment of an efficient student model while maintaining high accuracy (99.60\%) comparable to its teacher model (99.59\%). This lightweight approach ensures adaptability in resource-constrained environments, such as IoT devices, surpassing traditional ML-based intrusion detection systems that often require extensive computational resources. Furthermore, integrating XAI techniques like LIME enhances interpretability, addressing deep learning model's \textit{black-box} nature, a crucial factor for cybersecurity applications \cite{n7}. The integration of feature unlearning secure data privacy, a critical aspect in modern AI systems, aligns with recent advancements in machine unlearning \cite{n8}.

Using confusion matrices and the LIME explanation, we can validate that feature unlearning is working properly. In fig-\ref{total_con}, once the \textit{flgs} feature is unlearned, the teacher’s false positive rate increases from 12,870 to 20,445, and the student's false positive rate rises from 14,044 to 20,441, indicating that the models have genuinely stopped relying on this feature and that the decision perimeter has shifted. After restoring the feature, both models revert exactly to their original confusion matrix behavior, confirming the reversibility of the unlearning step. LIME further supports unlearning validation at the instance level, for test case-100, the student model initially has 8\% influence on the \textit{flgs} feature (fig-\ref{fig10}), but after unlearning, the flag feature no longer appears as a contributing factor for the same case (fig-\ref{fig12}).

Compared to benchmark models such as ensemble classifiers (e.g., LightGBM) and deep learning approaches like LSTM and RNN, our framework delivers competitive performance while placing stronger emphasis on data privacy through feature unlearning. Unlike prior studies that depend heavily on conventional feature engineering or lack robust privacy-preserving mechanisms, this work achieves a balanced trade-off between detection accuracy and regulatory compliance. Furthermore, by leveraging the BoT-IoT dataset, the evaluation reflects realistic cyberattack scenarios, reinforcing the framework’s practical applicability.

The study has limitations that aim for further research. The model primarily focuses on the BoT-IoT dataset, which, while comprehensive, may not capture arising attack variations. Future work could extend its applicability to diverse datasets, ensuring more comprehensive generalizability. Refining advance attention based reinforcement unlearning methods to remove sensitive data dynamically in real time could improve adaptability in evolving threat landscapes. Enhancing the model’s resilience against adversarial attacks through robust feature selection and hybrid learning strategies remains another critical avenue for exploration \cite{n6}.

\subsection{Discussion}
To further check the robustness of our framework, we tested the overall framework with 30\% of the data instead of just the 25\% used earlier. As shown in table-\ref{table_result}, the teacher and student models performed very well even with the extra data (added 5 million data points). The accuracy, F1 score, and loss values stayed consistently strong at each stage, which includes before feature removal, after feature removal, and after restoring the feature. This shows that the models are accurate and scalable even with big data. For example, with 30\% of the data before feature removal, the teacher model reached 99.471\% accuracy and a 99.735\% F1 score, slightly higher than 25\% of the data. Even after removing or restoring the feature, the models gave highly accurate results. These findings confirm that the framework remains effective and scalable.

To prove the computational efficiency of our proposed DiRLU model, we compared its FLOPS and parameter count with those of several well-known benchmark models used for botnet detection. DiRLU (student model) requires only 2,370 FLOPS, which is approximately 3.87$\times$ lower than KronNet’s 9,176 FLOPS on the same Bot-IoT dataset. In terms of parameter count, DiRLU has 3,014 parameters, which is very low. Compared with other benchmark models, DiRLU consistently shows substantially lower computational cost than DL-BiLSTM and IRNet-MBSKD. These results indicate that DiRLU is significantly more efficient and well-suited for deployment in resource-constrained IoT environments. The complete comparison is summarized in table-\ref{table_flops_params}.

\begin{table}[!ht]
    \centering
    \renewcommand{\arraystretch}{1.15}
    \setlength{\tabcolsep}{5pt}
    \caption{Efficiency comparison of DiRLU with benchmark models}

    \resizebox{0.95\linewidth}{!}{%
    \begin{tabularx}{\linewidth}{
        >{\raggedright\arraybackslash}p{2.35cm}
        >{\centering\arraybackslash}p{3cm}
        >{\centering\arraybackslash}p{1cm}
        >{\centering\arraybackslash}p{1cm}
    }
    \hline
    \textbf{Model} & \textbf{Dataset} & \textbf{FLOPS} & \textbf{Params} \\
    \hline

    DiRLU (student) & Bot-IoT & 2,370 & 3,014 \\ 
    KronNet \cite{KronNet} & Bot-IoT & 9,176 & N/A \\ 
    DL-BiLSTM \cite{DL-BiLSTM} & CIC-IDS2017, N-BaIoT \& CIC-IoT2023 & 628,800 & N/A \\ 
    IRNet-MBSKD \cite{IRNet-MBSKD} & NSL-KDD \& CIC-IDS2017 & 198.65K & 5.07K \\ 
    ConvNeXt-Sf \cite{ConvNeXt-Sf} & TON-IoT \& BoT-IoT & N/A & 0.26M \\ 
    DFF-FL \cite{DFF-FL} & CAN-Hacking & N/A & 81,863 \\
    \hline
    \end{tabularx}
    }

    \label{table_flops_params}
\end{table}

In this study, we faced many challenges, among which highly imbalanced data distribution was one of the significant. The imbalanced data distribution made it especially challenging to maintain high recall and precision for the benign class, which contained far fewer samples than the attack classes. Nevertheless, our framework achieved consistently strong performance across all evaluation metrics, demonstrating both robustness and adaptability. Feature analysis revealed that the state feature was particularly dominant, exerting greater influence than other features and posing a risk of bias if not properly managed. One of the framework’s main strengths lies in its lightweight and scalable design, enabling efficient training and deployment even in resource-constrained environments. In addition, the ability to apply post-hoc feature unlearning without retraining gives it a distinct advantage over conventional models. Unlike prior work, Cer-FeaUn \cite{CerFeaUn} has demonstrated certified feature unlearning in federated learning, while VERIFI \cite{VERIFI} has addressed the challenge of federated unlearning by enabling the removal of a participant’s contribution from a federated model. In contrast, this paper applies unlearning to IoT botnet detection through distilled reinforcement learning with explainability. Together, these capabilities allowed our approach to outperform several existing network security models in terms of accuracy, flexibility, and privacy-awareness. Overall, the results confirm the practicality and effectiveness of the framework for real-world, dynamic IoT security scenarios.

\section{Conclusion}\label{conclusion}
In this research, we introduced DiRLU, an optimized reinforcement learning-unlearning framework developed for IoT security. This framework includes knowledge distillation, which lets a compact student model learn from a larger teacher model. It also has a feature unlearning mechanism, that keeps sensitive data protected without having to retrain. Therefore, we utilized a subset of 25\% from the BoT-IoT dataset in comparison to the previous benchmark models that used only 5\%. Thus, our model achieved an accuracy of 99.60\% and an F1-score of 99.80\%. In this line, DiRLU is very fast as far as computation is concerned since it needs only 2370 FLOPS, which is approximately 3.87 times more efficient than the best model, KronNet. Our experiments show that with feature un-learning, it is possible to practically reverse the process – reducing sensitive features reduced the reliance on them while preserving performance, and restoring those features led to a full recovery of the original results. The incorporation of explainable AI made things clearer by providing a detailed scenario of how such decisions are made by the model system. All in all, DiRLU emerged quite beneficially and flexibly for the security of IoT devices. It combines efficiency, privacy protection, and ease of understanding to make it ready for use in the real world. Future work will concentrate on enhancing adaptability to new attack vectors and strengthening defenses against adversarial methods to guarantee secure and reliable IoT systems.

\section*{Code Availability}
The source code is available at the following hyperlink: \underline{\url{https://github.com/Nahidhasan07/Botnet-Traffic-Detection}}



\bibliographystyle{IEEEtran}
\bibliography{references.bib}

\end{document}